\documentclass[twocolumn,floatfix]{aastex63}

\usepackage{graphicx}	
\usepackage{amsmath}	
\usepackage{amssymb}	
\usepackage{bm}		
\usepackage{multirow}
\usepackage{dashrule}

\usepackage{placeins}

\received{xx}
\revised{xx}
\accepted{xx}

\def\apj{ApJ}
\def\mnras{MNRAS}
\def\nat{Nat}

\def\araa{ARA\&A}
\def\aap{A\&A}

\def\aj{AJ}
\def\apjs{ApJS}

\def\apjl{ApJ}

\def\nphysa{Nucl. Phys. A}

\def\prd{Phys. Rev. D}

\def\simge{\mathrel{\rlap{\raise 0.511ex
     \hbox{$>$}}{\lower 0.511ex \hbox{$\sim$}}}}
\def\simle{\mathrel{\rlap{\raise 0.511ex
      \hbox{$<$}}{\lower 0.511ex \hbox{$\sim$}}}}

\newcommand{\be}{\begin{equation}}
\newcommand{\ee}{\end{equation}}
\newcommand{\ba}{\begin{eqnarray}}
\newcommand{\ea}{\end{eqnarray}}

\newcommand{\eq}[1]{Eq.~\eqref{#1}}
\renewcommand{\fig}[1]{Fig.~\ref{#1}}
\newcommand{\tab}[1]{Table~\ref{#1}}
\newcommand{\sect}[1]{Section~\ref{#1}}
\newcommand{\app}[1]{Appendix~\ref{#1}}

\newcommand{\gcc}{\mathrm{~g \; cm}^{-3}}
\newcommand{\gcs}{\mathrm{g \; cm}^{-2}}
\newcommand{\ergs}{\mathrm{erg \; s}^{-1}}

\newcommand{\Lcs}{\mbox{$L_\mathrm{cs}\,$}}
\newcommand{\Lth}{\mbox{$L_\mathrm{th}\,$}}

\shorttitle{NS 1987A in SN 1987A}
\shortauthors{Page et al.}

\begin{document}

\title{NS 1987A in SN 1987A}

\author[0000-0003-2498-4326]{Dany Page}
\email{page@astro.unam.mx}
\affiliation{Instituto de Astronom\'ia, Universidad Nacional Aut\'onoma de M\'exico, Ciudad de M\'exico, CDMX 04510, Mexico}

\author[0000-0002-7326-7270]{Mikhail V. Beznogov}
\email{mikhail@astro.unam.mx}
\affiliation{Instituto de Astronom\'ia, Universidad Nacional Aut\'onoma de M\'exico, Ciudad de M\'exico, CDMX 04510, Mexico}

\author[0000-0001-7422-9690]{Iv\'an Garibay}
\email{igaribay@astro.unam.mx}
\affiliation{Instituto de Astronom\'ia, Universidad Nacional Aut\'onoma de M\'exico, Ciudad de M\'exico, CDMX 04510, Mexico}

\author[0000-0002-5907-4552]{James M. Lattimer}
\email{james.lattimer@stonybrook.edu}
\affiliation{Department of Physics and Astronomy, Stony Brook University, Stony Brook, NY 11794-3800, USA}

\author[0000-0002-9019-5029]{Madappa Prakash}
\email{prakash@ohio.edu}
\affiliation{Department of Physics and Astronomy, Ohio University, Athens, OH 45701, USA}

\author[0000-0002-0831-3330]{Hans-Thomas Janka}
\email{thj@mpa-garching.mpg.de}
\affiliation{Max-Planck-Institut f\"ur Astrophysik, P.O. Box 1317,
  Karl-Schwarzschild-Str. 1, 85741 Garching, Germany}

\begin{abstract}
The possible detection of a compact object  in the remnant of SN 1987A presents an unprecedented opportunity to follow its early evolution.
The suspected detection stems from an excess of infrared emission from a dust blob near the compact object's predicted position.  
The infrared excess could be due to the decay of  isotopes like $^{44}$Ti, accretion luminosity from a neutron star or black hole, magnetospheric emission or a wind originating from the spindown of a pulsar, or thermal emission from an embedded, cooling neutron star (NS 1987A).
It is shown that the last possibility is the most plausible as the other explanations are disfavored by other observations and/or require fine-tuning of parameters.  
Not only are there indications the dust blob overlaps the predicted location of a kicked compact remnant, but its excess luminosity also matches the expected thermal power of a 30 year old neutron star. 
Furthermore, models of cooling neutron stars within the Minimal Cooling paradigm readily fit both NS 1987A and Cas A, the next-youngest known neutron star.
If correct, a long  heat transport timescale in the crust and a large effective stellar temperature are favored, implying relatively limited crustal n-$^1$S$_0$ superfluidity and an envelope with a thick layer of light elements, respectively.
If the locations don't overlap, then  pulsar spindown or accretion might be more likely, but the pulsar's period and magnetic field or the accretion rate must be rather finely tuned. In this case, NS 1987A may have enhanced cooling and/or a heavy-element envelope. 
\end{abstract}


\section{Introduction} 
\label{sec:intro}
 
The chance to witness the birth of a neutron star arose when neutrinos were detected from the core-collapse supernova SN 1987A on Feb 23, 1987~\citep{Hirata:1987aa,Bionta:1987aa}.  
The further opportunity to observe the early evolution of a neutron star has, however, been elusive. 
Owing to the dust and the ring surrounding the supernova remnant (SNR) of SN 1987A, observational searches in radio and X-rays for a neutron star remnant have been unsuccessful \citep[e.g.,][]{Alp:2018ao,Esposito+2018}.  
Recently, however, \citet{Cigan:2019aa} have presented high angular resolution images of dust and molecules in  SN 1987A ejecta obtained from the Atacama Large Millimeter/submillimeter Array (ALMA) and concluded that the presence of a compact source in the remnant is strongly indicated.  
They observed a localized blob of warm dust with temperature $\simeq 33$ K 
compared to that of the surrounding dust, $\simeq (17$--$22)$ K.  
The observed SiO and CO gas temperatures correspond to a luminosity of the 
dust blob of \Lcs $= (40$--$90) L_\odot$, where $L_\odot = 3.826 \times 
10^{33}\,\ergs$ is the solar luminosity (the subscript `cs' stands for `condensed 
source'), which requires a compact source to have
an estimated power of this magnitude or slightly lower, if the source 
is embedded in the dust blob. However, if the compact source is not located
within the blob, but heats it from afar, the source must have a power somewhat greater than $\Lcs$.

In proposing a compact object in the remnant of SN 1987A as the most likely explanation for the observed excess dust blob emission, \citet{Cigan:2019aa} had to rule out alternative possibilities, the foremost of which was heating by the decay of $^{44}$Ti synthesized by SN 1987A. 
While the warm, extended dust is expected to be radioactively heated by $^{44}$Ti, this is unlikely for the concentrated, warmer blob. 
The existence of a single, high-density clump of $^{44}$Ti seems implausible, and even if it were formed, its heating would not be strongly localized because of optically thin conditions to the $^{44}$Ti $\gamma$-rays as discussed in \sect{Sec:ti-heating}.

\cite{Cigan:2019aa} {also} noted that there is an offset between the location of the brightest pixel of the warm blob and the center of the SNR at the original position of the progenitor star. 
This displacement, which could be associated with the supernova kick imparted to the compact object, is between about 20 mas and 85 mas, depending on how the center of the explosion is determined, {\it e.g.,} by fitting the geometrical center of the 315 GHz emission seen by ALMA \citep{Cigan:2019aa}, the ATCA radio ring continuum \citep{Potter:2009}, or the ring hot spots on HST images \citep{Alp:2018ao}.  
If connected with a supernova kick, the velocity component transverse to the line of sight is between 160 km/s and nearly 700 km/s (for the 51.4 kpc distance to the LMC).   
In addition, the kick, judging from the distribution of iron-group and intermediate-mass elements \citep{Larsson:2016}, should have a northerly component  in the sky \citep{Janka:2017aa}, which matches the orientation of the dust blob with respect to the original position of the progenitor. 
This offers evidence that the compact source is nearby or even surrounded by the blob. 

Specifically, \cite{Janka:2017aa} analysed the geometry of Fe and Si in a set of 3D supernova simulations for matching the shape and mass of the Fe+Si distribution of SN 1987A as determined by \cite{Larsson:2016}.
The best-fit model, L15-1, had also been considered \citep{Abellan:2017} with respect to the distribution of molecular CO 2-1 and SiO 5-4 emission in the ejecta of SN 1987A, and well-fits the size, shape and clumpy character of its apparent ring geometry.  
By orienting the Fe+Si ejecta of this model with respect to the ring plane and observer direction to obtain the asymmetry seen in SN 1987A, the supernova kick turned out to have a northern component. 
The main reason for this is a big Fe+Si mass located below the ring plane of SN 1987A, south of the connecting line to the observer. 3D explosion simulations show that the supernova kick vector and the bulk mass of iron-group and intermediate-mass elements should lie in opposite hemispheres 
\citep{Wongwathanarat:2013tu}, compatible with observations \citep{Katsuda:2018,Holland-Ashford:2017ck}.
The same 3D supernova model, L15-1, also allowed for a reasonable match of the redshift of the $^{44}$Ti emission and of the $^{56}$Co 847 keV line profile observed in SN 1987A \citep{Jerkstrand2020}.
The resulting angle between the supernova kick vector and observer direction should be about 30 degrees, and most likely less than about 90 degrees \citep{Jerkstrand2020}.  
The model L15-1 had a kick velocity only around 300 km/s, while the actual kick may have been 500 km/s or more, but the additional expense of fine-tuning was considered not worthwhile.

That the explosion was significantly asymmetric is not in doubt,  since radioactive $^{56}$Co debris is seen mostly moving away from us~ \citep{McCray:2016aa}. 
A NuSTAR observation \citep{Boggs:2015aa} also shows that $^{44}$Ti is considerably redshifted, suggesting a kick velocity component  along the line of sight towards us of several hundred km/s. 
With a transverse component of 160 km/s the compact source in SN1987A would have a space velocity near the peak of the distribution observed for young pulsars, whereas a transverse component of nearly 700 km/s would place it in the high-velocity tail.

Explosion models of \citet{Utrobin:2019} for state-of-the-art progenitor 
models of SN 1987A indicate the baryon mass, $M_B$, of its compact remnant 
to be $(1.35$--$1.66)\,M_\odot$, while \citet{Ertl:2019} predict 
$(1.48$--$1.56)\,M_\odot$ for single-star progenitors and 
$(1.38$--$1.75)\,M_\odot$ for binary progenitor\footnote{We have dropped 
results with the progenitors B15 in \citet{Utrobin:2019} and W15 in 
\cite{Ertl:2019} (which is the same model) because its He core mass is too 
small to explain the light curve peak, its pre-SN luminosity is too small, and it 
ejects too little O. Additionally, we omitted cases from \cite{Utrobin:2019} 
with too-little ejected nickel or too-high explosion energies.}. These baryon 
masses translate to a gravitational mass $M\simeq (1.22$--$1.62)\,M_\odot$ 
using the EOS-independent relation~\citep{Lattimer:2001}
\begin{equation}
\frac{M_B-M}{M} \simeq (1.2\pm0.1)\frac{\beta}{2-\beta} \,,
\label{eq:BGmasses}
\end{equation}
where $\beta=GM/Rc^2$ and $R\simeq11.5\pm1$ km is the typical neutron 
star radius\footnote{All radii quoted in this paper are circumferential radii, 
{\em i.e.}, no red-shift factor is applied. \label{footnote1}}.
These values are well below the measured masses, $M\simge 2\,M_\odot$, of 
several pulsars (PSR J1614-2230, \citealt{Demorest:2010aa}; PSR 
J0348+0432, \citealt{Antoniadis:2013aa}; and PSR J0740+6620, 
\citealt{Cromartie:2020aa}), as well as an inferred upper limit to the neutron 
star maximum mass $M_\mathrm{max}\simle (2.2$--$2.3)\,M_\odot$ 
\citep{Margalit:2017} from GW170817, which strongly suggests that a black 
hole
remnant in SN 1987A is unlikely.  
The identification of the compact remnant with an intermediate-mass neutron 
star is supported  by the observations of neutrinos from SN 1987A and the 
association of their inferred total energy \citep{Lattimer1989} with the binding 
energy of a proto-neutron star.  If about 1/6 of the total energy was radiated as 
the observed electron anti-neutrinos, the implied $(2.9\pm1.2)\times10^{53}$ 
ergs binding energy suggests the neutron star gravitational mass to be 
$(1.38\pm0.43)\,M_\odot$ using \eq{eq:BGmasses}. We assume in this paper 
that the compact remnant  produced by SN 1987A is most likely a neutron star, 
hereafter called NS 1987A, which is also possibly a pulsar.

If the neutron star is enclosed within the blob, the most natural explanation is that the blob is heated by its thermal emission \Lth.  
As we show in  \sect{Sec:minimal}, the expected \Lth of a 30 year old neutron star is within a factor 3 of the inferred excess blob luminosity.

An alternative explanation, that the entire spindown power of a pulsar heats the surrounding blob, is disfavored by the fine-tuning of the rotational period $P$ and magnetic field $B$ of the young neutron star that would be required.  Both $P$ and $B$ could have values up to 2 orders of magnitude higher or lower than what is necessary. The same argument would apply for another alternative, an accretion power source from a neutron star or black hole.  Here, a reasonable upper limit is $L_\mathrm{acc}<1.3(M/M_\odot)\cdot10^{39}$ erg s$^{-1}$, which is 10 times the Eddington luminosity.  But the lower limit is arbitrarily small, so the large range of possible accretion rates implies fine-tuning is required to obtain $L_\mathrm{acc}\sim\Lcs$.

However, it is also possible that the blob and the neutron star's locations are disjoint, a situation well known from the Crab Nebula, where the brightest part of the pulsar wind and the pulsar are spatially separated \citep{Weisskopf:2000,Gomez:2012}.  
In this case, the pulsar wind or  accretion explanation might be preferred since only a fraction of a source's power would be required.  
Either could easily be large enough (with plausible $P$ and $B$, or accretion rates) while the expected \Lth would be quite insufficient. 
But it should be noted that past observations \citep{Alp:2018ao} have set  
upper limits to the total (bolometric) emission of any kind of compact source 
of about $138\,L_\odot$ in the presence of dust, and $22\,L_\odot$ without 
dust.  
Even with dust, this is 1000 times smaller than the Crab's luminosity.  
An otherwise hidden pulsar or accretion source can thus have a luminosity at 
most $1.5\,\Lcs$--\,$3.5\,\,\Lcs$, so it must be unnaturally close to the blob and 
again 
raises the prospect of fine-tuning that would disfavor these hypotheses.
\sect{Sec:pulsar} considers the possibility that the required energy stems from the spin-down of a young pulsar. 

Our preferred hypothesis is, instead, that the power source of the blob is NS 1987A, a central compact object (CCO: \citealt{Pavlov:2002aa}), defined to be a young neutron star in a SNR whose luminosity, {$L_\mathrm{th}$,} is predominantly due to surface thermal emission.
\sect{Sec:minimal} examines the cooling of a star following the ``Minimal Cooling'' scenario~\citep{Page:2004fr,Page:2009dd,Page:2011aa} which assumes the lack of rapid neutrino cooling due to a direct Urca process \citep{Lattimer:1991aa}, emphasizing the importance of the envelope's chemical composition.
The question of whether or not light elements can survive in the envelope during the hot, early stages of a neutron star's life is addressed in \sect{Sec:burning}.  
\sect{Sec:DUrca} considers the case of a neutron star that has enhanced neutrino cooling, possibly because it is relatively massive. 
In \sect{Sec:Comps}, a comparison of the cooling trajectories of NS 1987A and the neutron star, Cas A, in the Cassiopeia A SNR is made.  
\sect{Sec:Disc} contains a discussion and  conclusions.
Essential details of the equation of state (EOS) models used in this work are given in \app{app:tech}.
The neutrino cooling processes considered are summarized in \app{app:DU}, and neutron superfluid gaps used in the inner crust are described in \app{app:SF}.

\section{Blob luminosity and $^{44}$Ti decay}
\label{Sec:ti-heating}

Radioactive decay of $^{44}$Ti might offer a possible explanation of the blob luminosity, which is $L_\mathrm{cs} = (1.5$--$3.5)\times 10^{35}$\,erg\,s$^{-1}$ \citep{Cigan:2019aa},  if $\gamma$-rays and positrons produced through the decay channel of
$^{44}\mathrm{Ti}\to ^{44}\mathrm{Sc} \\ \to ^{44}\mathrm{Ca}$ were efficiently thermalized in the blob medium. 
The decays of $^{44}$Ti to $^{44}$Sc proceed by electron capture, and the transition of $^{44}$Sc to $^{44}$Ca is almost exclusively by $\beta^+$ decays.
A corresponding upper limit to the luminosity is obtained for complete thermalization and given by
\begin{equation}
L_\mathrm{blob}(t) \le \left|\frac{\mathrm{d}N_\mathrm{Ti}(t)}{\mathrm{d}t}\right|\,
E_\mathrm{decay} = 
\frac{N_\mathrm{Ti}(t)}{\tau_0}\,E_\mathrm{decay} \,.
\label{eq:lblob}
\end{equation}
Above, $N_\mathrm{Ti}(t) = N_{\mathrm{Ti},0}\exp(-t/\tau_0)$ is the time-dependent number of $^{44}$Ti nuclei in the blob with $N_{\mathrm{Ti},0}$ being the initial number and $\tau_0\approx 85$\,yr their decay time. 
The energy release per $^{44}$Ti decay, $E_\mathrm{decay} \approx 2.9$\,MeV includes the 0.068\,MeV, 0.078\,MeV, and 1.157\,MeV $\gamma$-photons from $^{44}$Sc and $^{44}$Ca de-excitation, as well as the  energy ($2m_ec^2+\langle E_{e^+}\rangle$) from $e^+e^-$-annihilation of the emitted positron, which possesses an average kinetic energy of $\langle E_{e^+}\rangle\sim$ 0.6\,MeV \citep{Cameron1999}. 
The blob luminosity measured about $t_\mathrm{m} \approx 30$\,yr after the explosion thus leads to a constraint on the initial $^{44}$Ti mass contained by the blob:
\begin{equation}
M^\mathrm{blob}_{\mathrm{Ti},0} = N_{\mathrm{Ti},0}\,m_\mathrm{Ti} \ge
\mathrm{e}^{t_\mathrm{m}/\tau_0}\,
\frac{L_\mathrm{blob}(t_\mathrm{m})\,\tau_0\,m_\mathrm{Ti}}{E_\mathrm{decay}}\,,
\label{eq:mtiblob}
\end{equation}
where $m_\mathrm{Ti}$ is the mass of a $^{44}$Ti atom.
Inserting numbers, including $L_\mathrm{blob}(t_\mathrm{m}) = \Lcs$,
\begin{equation}
M^\mathrm{blob}_{\mathrm{Ti},0} \ge (4.5-10.5)\times 10^{-6}\,M_\odot,
\label{eq:mtiblob2}
\end{equation}
which accounts for $\sim$3\%--7\% of the total mass of $^{44}$Ti ejected in 
SN~1987A ($\sim1.5\times 10^{-4}$\,$M_\odot$) as deduced from X-ray 
observations and light-curve analysis \citep{Boggs:2015aa,Jerkstrand2011}.

Mixing processes in 3D SN explosion models yield an initial $^{44}$Ti mass fraction relative to the freshly produced $^{56}$Ni mass fraction that varies within the interval 0.001--0.004 on large scales \citep{Wongwathanarat2017}. Therefore, one estimates the initial mass ratio of radioactive titanium and nickel in the blob to be
\begin{equation}
10^{-3} \lesssim {M^\mathrm{blob}_{\mathrm{Ti},0}}/{M^\mathrm{blob}_{\mathrm{Ni},0}}
\lesssim 4\times 10^{-3} \,,
\label{eq:massrat}
\end{equation}
and the corresponding mass of $^{56}$Ni in the blob to be
\begin{equation}
250\,M^\mathrm{blob}_{\mathrm{Ti},0} \lesssim M^\mathrm{blob}_{\mathrm{Ni},0}
\lesssim 1000\,M^\mathrm{blob}_{\mathrm{Ti},0} \,.
\label{eq:mniblob}
\end{equation}
With Eq.~(\ref{eq:mtiblob2}), one thus derives
\begin{eqnarray}
(1.1-2.6)\!\times\! 10^{-3}\,\mathrm{M}_\odot &\lesssim&
M^\mathrm{blob}_{\mathrm{Ni},0} \nonumber \\
&\lesssim&
(4.5-10.5)\!\times\!10^{-3}\,M_\odot,
\label{eq:mniblob2}
\end{eqnarray}
which means that the blob should carry between 1.6\% and 14.6\% of the total 
$^{56}$Ni mass (0.072\,$M_\odot$) produced by SN~1987A.

The lower limit of this range is in the ballpark of the radioactive nickel mass, 
about $10^{-3}$\,$M_\odot$, that has been estimated to be connected to a 
feature of the H$\alpha$ line profile during the so-called Bochum event 
\citep{Utrobin1995}. \citet{Wang2002} discussed an association of the 
$^{56}$Ni clumps of the Bochum event with the distribution of $^{44}$Ti 
that powers the late-time ejecta.
They argued that both share the same orientation with respect to the line of sight, namely a location north of it and away from the observer (i.e., redshifted, consistent with the $^{44}$Ti decay-line measurements by \citealt{Boggs:2015aa}) and along the symmetry axis of the ejecta that is inclined to the east by about 14$^\circ$ relative to the symmetry axis of the ring. 
This inferred apparent position of the $^{56}$Ni clumps and $^{44}$Ti heating north-east of the line of sight adds relevance to the question of whether the blob emission is powered by $^{44}$Ti decay.

However, in the following we will argue that this possibility is disfavored by the fact that the blob is not massive enough to enable efficient thermalization of the $\gamma$ photons produced by radioactive decays. 
The optical depth of the blob is of the order of
\begin{equation}
\tau_\gamma \sim \int_0^{R_\mathrm{blob}} \mathrm{d}r\,\kappa_\gamma\,\rho_\mathrm{blob}\,,
\label{eq:optdepth}
\end{equation}
where $R_\mathrm{blob}$ and $\rho_\mathrm{blob}$ are the blob radius and average density, respectively.
For radiation particles $i$, the opacity is $\kappa_i = Y_e\,m_\mathrm{b}^{-1}\sigma_i$, with $Y_e$ being the electron faction, $m_\mathrm{b}$ the average baryon mass, and $\sigma_i$ the reaction cross section.
The cross section is given by the Klein-Nishina formula for standard Compton scattering of the 68\,keV, 78\,keV, and 1157\,keV $\gamma$ photons, which are $\sigma_{68} \approx \sigma_{78} = 0.8\sigma_\mathrm{T}$ and $\sigma_{1157} = 0.3 \sigma_\mathrm{T}$, respectively, where $\sigma_\mathrm{T} \approx 6.65\times 10^{-25}$\,cm$^2$ is the Thomson scattering cross section of electromagnetic radiation with electrons.
The $^{44}$Sc decay positrons possess a broad energy spectrum with a peak around 600\,keV and a maximum energy of 1474\,keV. 
Their typical annihilation cross section with electrons is of the order $\lesssim\sigma_\mathrm{T}$, but the $\gamma$-rays emitted by the annihilation have a Klein-Nishina cross section that is smaller, namely in the range of 
$\sim$(0.29--0.43)$\sigma_\mathrm{T}$. 
A safe upper limit of the $\gamma$ opacity of the blob is therefore obtained with $\sigma_\gamma \le 0.8\sigma_\mathrm{T}$, and is given by
\begin{equation}
\kappa_\gamma \lesssim 0.16\,\left(\frac{Y_e}{0.5}\right)\ \mathrm{\frac{cm^2}{g}}\,.
\label{eq:opacity}
\end{equation}
The 5$\sigma$ emission knot, the blob, in \citet{Cigan:2019aa} has an angular diameter of $\theta_\mathrm{blob}\approx 20$\,mas. With a distance of 51.4\,kpc this converts to a blob radius
\begin{equation}
R_\mathrm{blob} \approx 7.66\times 10^{15}\,
\left(\frac{\theta_\mathrm{blob}}{20\,\mathrm{mas}}\right)\ \mathrm{cm}\,.
\label{eq:rblob}
\end{equation}
The average blob density can be expressed as
\begin{equation}
\rho_\mathrm{blob} \approx \frac{3}{4\pi}\,
\frac{M_\mathrm{blob}}{R_\mathrm{blob}^3} = 
\frac{3}{4\pi}\,\frac{M^\mathrm{blob}_\mathrm{Ni,0}}{R_\mathrm{blob}^3
X^\mathrm{blob}_\mathrm{Ni,0}} \,,
\label{eq:rhoblob}
\end{equation}
where $X^\mathrm{blob}_\mathrm{Ni,0}$ is the initial mass fraction of $^{56}$Ni in the clump of ejecta forming the blob. 
Using Eqs.~(\ref{eq:opacity})--(\ref{eq:rhoblob}) in Eq.~(\ref{eq:optdepth}) leads to
\begin{eqnarray}
\tau_\gamma &\sim& \kappa_\gamma\,R_\mathrm{blob}\,\rho_\mathrm{blob} \nonumber \\
&\lesssim&
1.3\times 10^{-2}\,
\left(\frac{M^\mathrm{blob}_\mathrm{Ni,0}}{0.01\,M_\odot}\right)
\left(\frac{\kappa_\gamma}{0.16\,\mathrm{cm}^2\,\mathrm{g}^{-1}}\right)\nonumber \\
&\phantom{\lesssim}&
\phantom{1.3\,10^{-2}}\,\,\times
(X^\mathrm{blob}_\mathrm{Ni,0})^{-1}
\left(\frac{\theta_\mathrm{blob}}{20\,\mathrm{mas}}\right)^{\!\!-2}\,,
\label{eq:taulimit}
\end{eqnarray}
where the normalization of $M^\mathrm{blob}_\mathrm{Ni,0}$ is guided by the upper limit of the mass interval given in Eq.~(\ref{eq:mniblob2}) and yields an upper limit of $\tau_\gamma$.
For efficient $\gamma$ thermalization it is necessary that  $\tau_\gamma \gtrsim 1$, which requires, at least, that
\begin{equation}
X^\mathrm{blob}_\mathrm{Ni,0} \lesssim 1.3\times 10^{-2}\,
\left(\frac{M^\mathrm{blob}_\mathrm{Ni,0}}{0.01\,M_\odot}\right)\,,
\label{eq:Xbloblimit}
\end{equation}
if all other factors in Eq.~(\ref{eq:taulimit}) are chosen to be unity.
With $X^\mathrm{blob}_\mathrm{Ni,0} = M^\mathrm{blob}_\mathrm{Ni,0}\,M_\mathrm{blob}^{-1}$, this equation leads to a lower limit of the blob mass:
\begin{equation}
M_\mathrm{blob} \gtrsim 0.77\,M_\odot \,.
\label{eq:Mbloblimit}
\end{equation}
Therefore, the condition $\tau_\gamma \gtrsim 1$ can only be fulfilled with an implausibly large mass of the blob, nearly as much as the entire mass of oxygen ejected in SN~1987A. 

For all reasonable values of $M_\mathrm{blob}$, we therefore conclude that $\tau_\gamma \ll 1$ and thermalization of the $\gamma$'s released by $^{44}$Ti decay in the blob is highly inefficient. 
Under such circumstances, with only partial thermalization of the $\gamma$ photons being possible, a $^{44}$Ti mass much higher than our estimate of Eq.~(\ref{eq:mtiblob2}) would be needed to explain $L_\mathrm{cs}$. 
Again, this is implausible and it would require extreme fine-tuning if several tens of percent of the ejected $^{44}$Ti, or even all of it, were concentrated in a single blob instead of being widely distributed in association with the expelled iron-group elements \citep[for corresponding 3D explosion simulations in comparison to the  gaseous remnant of Cas~A, see][]{Wongwathanarat2017}. 
For all these reasons $^{44}$Ti-decay heating cannot provide a convincing explanation of the entire blob luminosity.

Nevertheless, a small fraction of the blob luminosity of (40--90)\,$L_\odot$
might still come from $^{44}$Ti-decay heating~\citep{Cigan:2019aa} or irradiation by external sources. If such mechanisms were responsible for heating the blob to the $\sim$22\,K of its surroundings, they would produce only a fraction of $(22/33)^4\approx 20$\,\% of the blob luminosity (blackbody emission is assumed for this estimate). An over-dense cloud of dust, which is not implausible for the blob due to the strong gravitational attraction of an enclosed compact object, might be heated by radioactivity even somewhat above the 22\,K of the surroundings. For a 25\,K radioactive ``floor value'' of the blob temperature, for example, the $^{44}$Ti-decay would account for $(25/33)^4\approx 33$\,\% of the blob luminosity.

\section{The Spin-down Power of Young Pulsars and Central Compact Objects} 
\label{Sec:pulsar}

\begin{figure}[t]
\includegraphics[width=1.10\columnwidth]{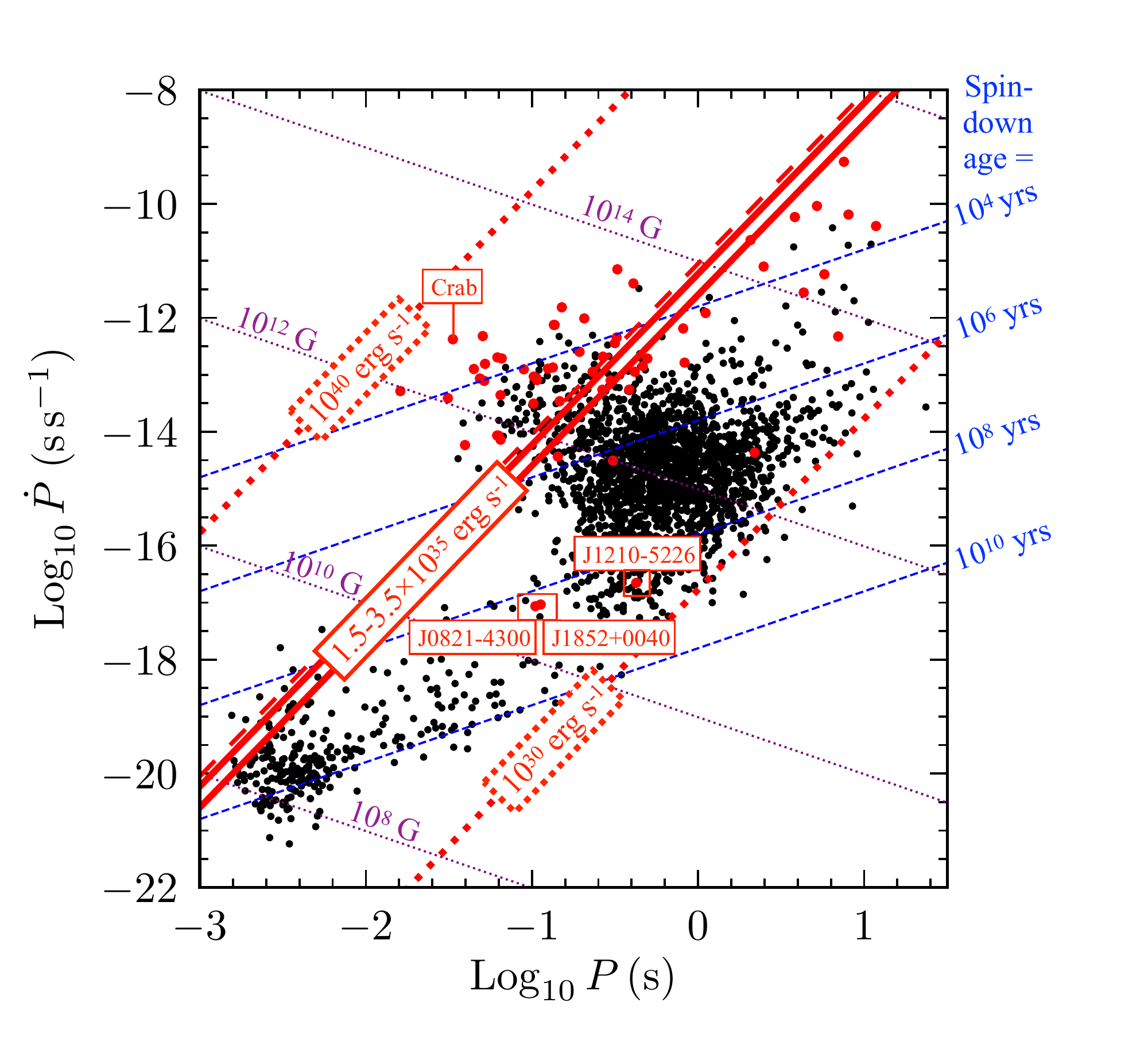}
\caption{Measured period derivatives $\dot P$ vs periods $P$ for 2,256 pulsars 
from the ATNF Pulsar Catalogue (\citealt{Manchester:2005aa}, and its on-line 
version at \url{https://www.atnf.csiro.au/research/pulsar/psrcat}).
Red dots show pulsars with a SNR association.
Shown are lines of constant spin-down timescale $\tau_\mathrm{sd} \equiv P/(2\dot P)$ (dashed blue),  inferred polar magnetic field strength $B_p \equiv 3.2 \times 10^{19} (P\dot P)^{1/2}$~G \citep{Lyne:2012aa} (violet dots), and spin-down power $W \equiv | I \Omega \dot \Omega|$ (red dots), where $\Omega \equiv 2\pi/P$ and an assumed moment of inertia $I = 1.5 \times 10^{45}$ g cm$^2$ \citep{Haensel:2007aa}.
The dark red lines delimit the inferred {range} $\Lcs\sim (40$--$90)\,L_\odot = 
(1.5$--$3.5) \times 10^{35}$ $\ergs$ corresponding to NS 1987A; the adjacent 
thin dashed red line shows the upper limit of $138\, L_\odot = 5.2\times 
10^{35}$ $\ergs$
for the luminosity of any compact source in the SNR \citep{Alp:2018ao}.
The locations of the Crab pulsar and the only three CCOs with measured $P$ and $\dot P$, PSR J1852+0040 in Kes 79 \citep{Halpern:2010aa}, PSR 0821-4300 in Puppis A, and PSR J1210-5226 in PKS 1209-52 \citep{Gotthelf:2013aa}, are indicated.}
\label{Fig:P_Pdot}
\end{figure}

An obvious explanation for the source of energy powering the dust blob is the {pulsar's} spin-down power $W \equiv | I \Omega \dot \Omega|$, where $I$ is the moment of inertia and $\Omega$ is the rotational frequency.
Fig.~\ref{Fig:P_Pdot} shows  the $P$-$\dot P$ diagram containing known 
pulsars to put this hypothesis in context.
While the {needed} luminosity \Lcs is within the range of values for young pulsars associated with a SNR, that range is very large, extending from $\sim 2\times 10^{31} \; \ergs$ in the case of PSR J1210-5226 up to $\simeq 10^{39} \; \ergs$  for the Crab pulsar.
Moreover, most young pulsars with ``canonical''  polar magnetic field strengths 
$B_p \sim 10^{12-13}$~G have $W\gg \Lcs$, while magnetar-type pulsars 
with $B_p > 10^{14}$ G exhibit values substantially below \Lcs .
The three CCOs with measured $P$ and $\dot P$, which must also be young pulsars and are displayed in \fig{Fig:P_Pdot}, also have $W\ll \Lcs$.

The large possible range for $W$ suggests fine-tuning is involved if a pulsar is embedded in the blob and powering its excess luminosity.  
It must lie between the double red lines in Fig. \ref{Fig:P_Pdot}.  
And it may be even more unlikely that the pulsar is located outside the blob, since in this case $W$ must be larger than \Lcs and the pulsar must lie above the double red lines in Fig. \ref{Fig:P_Pdot}. 
But there is an observed \citep{Alp:2018ao} upper limit to the power of a 
hidden pulsar in the 1987A SNR that is only $138L_\odot$, i.e., $\simeq 
(1.5$--$3.5)\,\Lcs$, suggesting the pulsar is located incredibly close to the blob.

The emerging family of CCOs comprises about a dozen objects, all young by 
definition as they are found in SNRs (see the on-line listing at 
\url{http://www.iasf-milano.inaf.it/~deluca/cco/main.htm}), and, except 
for the three mentioned above, generally presenting no evidence for a 
significant magnetic field (see, e.g., 
\citealt{Gotthelf:2008lq,de-Luca:2008cz,De-Luca:2017wg} for reviews).
The ATNF Pulsar Catalogue \citep{Manchester:2005aa} lists 32 pulsars with spin-down timescales $\tau_\mathrm{sd} \equiv P/(2\dot P)<10,000$ yrs.
Of these, 21 are ``regular'' pulsars with $B_p < 10^{14}$ G, and 11 are ``magnetars'', i.e., having $B_p > 10^{14}$ G.
Since 12 CCOs are known, this indicates that a young neutron star has a significant $12/(21+11+12)\simeq27\%$ chance of being observed as a CCO.  
(Of course, the actual probability of being a CCO could be different because of selection biases.)
Indeed, NS 1987A had been proposed to be a CCO before the class was even given a name \citep{Muslimov:1995po,Page:1998ip,Geppert:1999tl}.

The reasonably large probability of being observed as a CCO and the fine-tuning in $P$ and $B_p$ required to explain the blob's luminosity whether or not it encloses the neutron star imply the energy source powering the dust blob is most likely of a different nature than spin-down.
We therefore favor the idea that NS 1987A is a CCO.
In the next section, it is shown that \Lth\!, except under certain circumstances, is naturally the same magnitude as \Lcs\!. 

\section{The Thermal Luminosity of a 30 Year Old Neutron Star} 
\label{Sec:minimal}

The early thermal evolution of a neutron star has been specifically studied in 
several works (see, e.g., 
\citealt{Lattimer:1994aa,Gnedin:2001aa,2008AstL...34..675S}) and in most 
cases the luminosity at age 30 years is within a factor 0.1--1 of \Lcs.
Here we examine the physical conditions  that control a young neutron star's luminosity around this age.

Subsequent to its birth in a core-collapse SN, the compact remnant enters the proto-neutron star phase in which  hot and dense matter temporarily traps neutrinos \citep{Burrows:1986aa}.
This phase lasts a minute or so due to eventual energy and lepton number loss from neutrino leakage after which matter reaches its beta-equilibrated chemical composition.
After this, the neo-neutron star \citep{Beznogov:2020aa} has a rapidly 
decreasing \Lth for about a year until reaching a plateau of $(0.5$--$5) \times 
10^{35}$ erg s$^{-1}$ ($12$--$120\,L_\odot$) that lasts for a few decades.
This range agrees well with \Lcs.

In the neo-neutron star phase, the surface temperature $T_\mathrm{s}$ is initially controlled mainly by the matter directly underneath it.  
With progressing time, however, conduction allows deeper and deeper layers to play a role.
The early cooling, after the first day, is entirely due to neutrino emission by plasmon decay ($\gamma^* \rightarrow \nu \bar \nu$) and results in a virtually universal temperature-density ($T$-$\rho$) profile in the outer layers.
When the local $T(\rho)$ has dropped below the local plasma temperature $T_P(\rho)$, the plasmon density is exponentially reduced, and, consequently, plasmon decay ceases as well \citep{Adams:1963oz}. 
Hence, the $T(\rho)$ profile follows from the electron plasma temperature $T_{P, \, e} \equiv \hbar \Omega_{P, \, e}/k_B$ where $\Omega_{P, \, e}^2 \equiv 4\pi e^2 n_e/m_e^*$ and $n_e$ is the net electron density.  
The effective mass of ultra-relativistic electrons is $m_e^* = \hbar k_{F}(e)/c$, where  the electron Fermi momentum $k_{F}(e)\equiv (3 \pi^2 n_e)^{1/3}$.  
Thus, $T(\rho)$ is directly determined by the $n_e(\rho)$ profile.
The standard assumption is that, being formed by cooling from very high temperatures, the crust of a neo-neutron star consists of ``catalyzed matter'' in which the optimum nucleus minimizes  the energy density \citep{Baym:1971aa} at each baryon number density.
At low-enough densities,  this implies the presence of $^{56}$Fe.
As a consequence, $n_e(\rho)$ is a universal function, and, from the above discussion and before thermal diffusion takes over in a very young star, so is $T(\rho)$ (see \citealt{Beznogov:2020aa} for more details).

Once plasmon decay shuts off, crustal neutrino emission continues from electron-ion bremsstrahlung, which is, however, relatively inefficient \citep{Gnedin:2001aa}. 
Hence, subsequent cooling proceeds very slowly with further evolution mainly determined by heat transport.
The core maintains more powerful neutrino emission and is therefore colder than the crust, resulting in the crustal thermal energy diffusing inward, which leads to the decrease of $T_\mathrm{s}$ on the thermal diffusion time scale.

Thermal diffusion through a layer of width $l$ acts on a timescale \citep{Brown:1988aa}
\begin{equation}
\tau_\mathrm{th} \approx l^2 \, {C_V}/{K} \,,
\label{Eq:tau}
\end{equation}
where $C_V$ is the specific heat per unit volume and $K$ is the thermal conductivity of the layer.
The width $l$ of any crustal layer scales with the total crust thickness $\Delta R$, which is primarily a function of the star's mass $M$ and radius $R$ \citep{Lattimer:1994aa}. 
The specific heat poses a major uncertainty due to the presence of dripped neutrons, which  contribute negligibly if they are in a superfluid state (i.e., if the local $T(\rho)<T_\mathrm{c}(\rho),$ the local superfluid transition critical temperature), whereas they will largely dominate if they are in the normal state. 
Uncertainty in knowledge of $T_\mathrm{c}(\rho)$ directly translates into a large uncertainty in $C_V$ and has a significant impact on the thermal relaxation of the crust as shown below.

In contrast, the thermal conductivity is dominated by well-understood electron scattering so that $K(\rho)$ in itself has little uncertainty once the chemical composition of the medium is determined\footnote{Electron scattering by impurities and lattice defaults is the major source of uncertainty but only at temperatures much lower than those present in the crust of a young neutron star. 
The reader is referred to \citet{Page:2012ys} for a review on the thermal properties of a neutron star's crust.}.
Catalyzed matter in the outer layers implies the presence of $^{56}$Fe up to densities $\sim 10^7 \; \gcc$ and increasingly more neutron-rich nuclei as density increases.
However, an important aspect of the problem is the possibility that some amount of light elements (H/He/C/O) may have been accreted  after the crust originally equilibrates.  
These light elements have smaller $Z$ and electron-ion scattering rates, and therefore larger thermal conductivities, than heavier elements, which lead to a smaller thermal gradient in the outer layers and higher surface temperature $T_\mathrm{s}$.
Thicker light-element layers correspond to larger $T_\mathrm{s}$ \citep{Potekhin:1997aa}.
We will call this outer layer the {\em  envelope,}\footnote{Traditionally, see, e.g., \citet{2006NuPhA.777..497P}, the envelope encompasses the outer layers at densities below a boundary value $\rho_\mathrm{b}$. 
These layers are treated separately for numerical convenience and {provide} the outer boundary condition as a ``$T_\mathrm{b}$-$T_\mathrm{s}$''relationship \citep{Gudmundsson:1983aa} between the boundary temperature $T_\mathrm{b}$ and the surface $T_\mathrm{s}$. 
Our present definition of {the} envelope is thus an extension to include all layers with possible non-catalyzed chemical composition extending to a density $\rho_\mathrm{c} = 10^{11} \gcc$ so that $\rho_{\cal{L}} \leq \rho_\mathrm{c}$ always.
We thus split the outer layers in an outer envelope at densities $\rho \leq \rho_\mathrm{b}$ included into the outer boundary condition, and an inner envelope with $\rho_\mathrm{b} \leq \rho \leq \rho_\mathrm{c}$ possibly contaminated by light elements (see \citealt{Beznogov:2020aa} for a similar extension).}
where  a strong temperature gradient toward the surface may be present and the chemical composition may have been altered from that of the original catalyzed matter.
We denote by $\rho_{\cal{L}}=10^{\cal{L}}$ g\,cm$^{-3}$ the maximum density reached by light elements, there being catalyzed matter at higher densities.

\begin{figure*}[t]
\centerline{\includegraphics[width=1.03\textwidth]{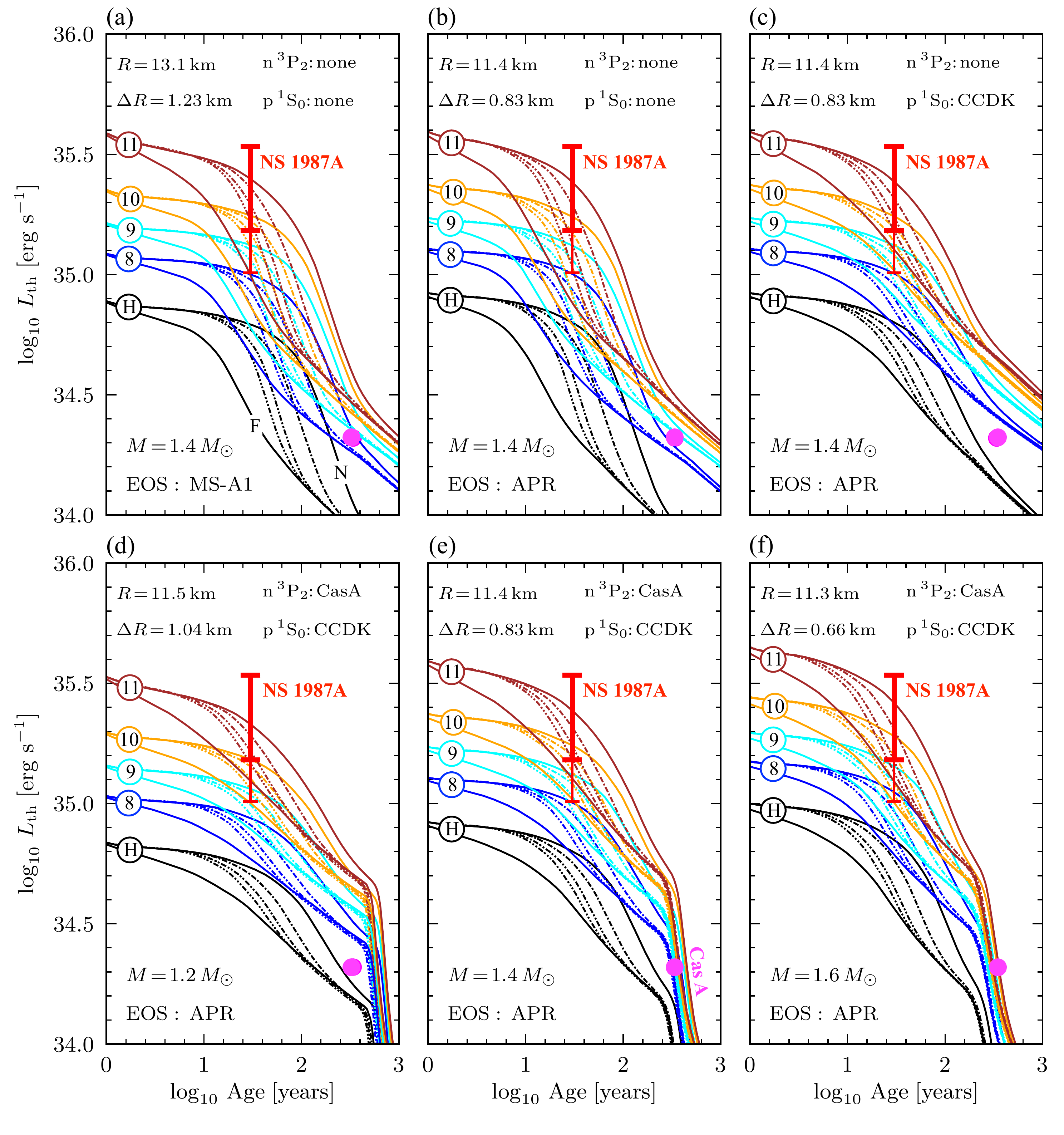}}
\caption{Evolution of the redshifted thermal luminosity \Lth for a young  neutron star.
Panel (a) models a stiff EOS, MS-A1; panels (b) through (f) model a soft EOS, APR (see \app{app:tech} for details).
The star's mass $M$, radius $R$ and crust thickness $\Delta R$ are  indicated.
Models with pure heavy element (Fe) envelopes are labeled H and those with increasing amounts of light elements (H, He, C, O) up to a maximum density $\rho_{\cal L}=10^{\cal L}$ are labeled with ${\cal L} =$ 8, 9, 10 and 11. The five curves within each of these 5 families represent different assumptions about the n-$^1$S$_0$ gap in the crust (see the text and \app{app:SF} for details).
Panels (a), (b) and (c) have no core n-$^3$P$_2$ superfluidity but d, (e) and (f) have this adjusted \citep{Page:2011aa,Shternin:2011aa} so that models with $\mathrm{log}_{10} \rho_{\cal L} = 8$ fit observed values of $\Lth$ and $d\Lth/dt$ of Cas A 
\citep{Heinke:2010aa} with the $1.4 M_\odot$ model. 
Panels (a) and (b) assume no core p-superconductivity and panels (c) through (f) have a large p-$^1$S$_0$ gap (model CCDK from \citet{Chen:1993xq}). 
The red error bar shows the age and inferred luminosity 
$\Lth=(40$--$90)\,L_\odot$ of NS~1987A (with the downward extension to 
$26\,L_\odot$ indicating that $\Lcs$ is an upper limit); the magenta dot 
denotes Cas~A's observed age and \Lth.}
\label{Fig:Cool_SF_CasA}
\end{figure*}

To quantify the above considerations, we have performed extensive numerical simulations using the code {\tt NSCool} \citep{Page:1989aa,Page:2016ff} that we previously used in \citet{Page:2004fr,Page:2009dd,Page:2011aa}. 
Our results are presented in \fig{Fig:Cool_SF_CasA}.
Each panel corresponds to different assumptions about the core's physics ({\it i.e.,} the dense matter EOS and core superfluidity)
or the stellar mass, whereas the various cooling curves in each panel correspond to different assumptions about the crust's physics (superfluidity) and the envelope's composition.

Although the early evolution is dominated by the crust and envelope, the neutron star core also plays a role through two different effects.
First, the core's EOS  determines the stellar mass $M$, radius $R$ and crust thickness $\Delta R$.
The latter determines the crustal length scales $l$ entering \eq{Eq:tau}.
Simulations are restricted to EOSs satisfying radius constraints described in \app{app:tech}. 
From the four EOSs we consider in \app{app:tech}, only results
for MS-A1 and APR that have the largest and smallest $R$ (and $\Delta R$), respectively, are displayed.  
The other two EOSs lead to intermediate results. 

Second, neutrino cooling of the core determines the final crust temperature once its thermal relaxation is completed.
Core cooling generally follows one of two paths, ``standard" (modified Urca or MU) or ``enhanced" (direct Urca or DU) cooling (see \app{app:DU}).
In this section, we only consider standard cooling, and, in \sect{Sec:DUrca}, we consider enhanced cooling which results in colder cores and possibly shorter crust relaxation time since $C_V$ decreases and $K$ increases (\eq{Eq:tau}).
If restricted to standard or slow neutrino cooling, {\it i.e.,} adhering to the Minimal Cooling paradigm, core neutrino emission is mostly controlled by nucleon pairing, either p-superconductivity or n-superfluidity.
The main effects of pairing are to suppress the dominant MU process and/or trigger the efficient Cooper pair breaking and formation processes (see, e.g., \citealt{Page:2014aa}).

Panels (a) and (b) of \fig{Fig:Cool_SF_CasA} explore the consequences of variations in $\Delta R$ due to the core's EOS under the assumption of no core pairing.  EOS-induced changes are seen to be small. 
Panels (b) and (c) explore the suppression of the MU emission by p-superconductivity, resulting in higher luminosities after crust relaxation and the consequent smaller temporal extension of the crust relaxation phase.
The assumed p-pairing model, CCDK from \citet{Chen:1993xq}, has extensive p-superconductivity covering almost the entire stellar core,  thus maximizing the suppression of neutrino cooling.
Finally, panels (d) through (f) display the effect of adding core n-$^3$P$_2$ 
superfluidity with a small gap as needed \citep{Page:2011aa,Shternin:2011aa} 
to explain the observed \Lth and $d\Lth/dt$ of the neutron star (Cas A) in the 
Cassiopeia A SNR assuming $M=1.4\,M_\odot$ \citep{Heinke:2010aa}. The 
influence of moderate variations in the stellar mass are explored in panels (d), 
(e) and (f).
The n-pairing phase transition is initiated in some core layer at age $\sim 250$ 
yrs (later for $1.2\,M_\odot$ and earlier for $1.6\,M_\odot$), resulting in a 
sudden increase in neutrino losses by the continuous formation and breaking 
of the n Cooper pairs and attendant sudden drop in \Lth.
Naturally, evolutions of \Lth at times before the onset of the superfluid phase transition in panel (e) are identical to those in panel (c).
A larger neutron gap in the core, resulting in an earlier onset of n-superfluidity, would have little consequence for NS 1987A and results in an evolution similar to the one in panel (c) \citep{Page:2009dd}.  
Although differing in some details, evolutions are insensitive to variations in 
the core's physics and stellar mass during the first 30--40 years of interest for 
NS 1987A.

Each panel of \fig{Fig:Cool_SF_CasA} contains five families of curves, each of which has 5 members. 
Each family represents a different envelope composition: the bottom-most family (H) representing pure heavy-element envelopes, followed by families ${\cal L}=$(8)--(11) having light elements present to increasingly higher maximum densities $\rho_{\cal L} = 10^{\cal L} \gcc$.  
The family members represent different assumptions about the crust's n-$^1$S$_0$ critical temperature $T_\mathrm{c}(\rho)$: from top to bottom there is a steady increase in the effective extent of crustal superfluidity at any time (see \app{app:SF}).

\fig{Fig:Cool_SF_CasA} highlights that, by far, the most important physical ingredient controlling the thermal luminosity at young ages is the chemical composition of the envelope.
This is evident from the strong increase of $L_\mathrm{th}$ from pure heavy-element envelopes to increasingly light-element-dominated envelopes after a few years.
Of secondary, but not negligible, importance is the extent of n-$^1$S$_0$ superfluidity in the inner crust.  
The differences are mostly due to the suppression of the neutron specific heat \citep{Page:2009dd}.
Models in which $T_\mathrm{c}$ grows more rapidly with increasing density in the inner crust have thicker layers of superfluid neutrons and cool faster, culminating with models that have fully superfluid (F) crusts.

Satisfactory agreement between the theoretical models and the inferred luminosity of NS 1987A requires not only relatively large amounts of light elements in the envelope, but also relatively small n-$^1$S$_0$ gaps near the neutron drip point, 
irrespective of values for $R$ and $\Delta R$.

The necessity of large amounts of light elements in the envelope could be 
relaxed somewhat (to ${\cal L}\sim (9)$) if the neutron star mass is relatively 
large. This can be seen in Fig.~\ref{Fig:Cool_SF_CasA}, where results for $M 
= 1.6\,M_\odot$ are also shown, i.e., for a neutron star mass near the upper end 
of the range that seems likely for SN~1987A. Moreover, the observationally 
inferred luminosity  $\Lcs=(40$--$90)\,L_\odot$ of the dust blob is only an 
upper limit to $L_\mathrm{th}$ of NS~1987A, because the radioactive decay 
of $^{44}$Ti, which already heats the dust in the blob surroundings to a 
temperature of $\sim$22\,K \citep{Cigan:2019aa}, provides additional heating. 
Correspondingly, the red error bar for $\Lth$ in Fig.~\ref{Fig:Cool_SF_CasA} 
stands for an upper limit,
and its downward extension indicates a reduction of 33\% to $26\,L_\odot$  
(see Sect.~\ref{Sec:ti-heating}).
This reduction also permits less massive light-element envelopes (${\cal L}\sim (8)$) to become compatible with NS 1987A.

One can relate the maximum density $\rho_{\cal L}$ reached by the light elements in the crust to the total mass of light elements through $M_{\cal L} = 4\pi R^2y_{\cal L}$, where $y_{\cal L}$ is the column density of the light element layer.  
The latter is determined by the pressure  $p_{\cal L} = y_{\cal L} g_\mathrm{s}$ at $\rho_{\cal L}$, 
where $g_\mathrm{s}=e^{-\phi}GM/R^2$ is the surface gravity ($e^\phi$ being the surface red-shift factor).
Since the pressure within the outer crust is dominated by ultra-relativistic degenerate electrons, one gets
\begin{eqnarray}
M_{\cal L} &=& 2.7 \times 10^{-6} \left(\frac{R}{12~{\rm km}}\right)^4\frac{M_\odot}{M} e^\phi\nonumber\\
&\times& \left(\frac{\rho_{\cal L}}{10^{10}\rm{~g~ cm}^{-3}}\frac{\langle Z \rangle}{\langle A\rangle}\right)^{4/3} M_\odot,
\end{eqnarray}
where $\langle Z \rangle$ and $\langle A \rangle$ are the average charge and mass numbers of the light element(s) present at $\rho_{\cal L}$.
$M_{\cal L}$ is thus a tiny fraction (at most, a hundredth of a percent) of the 
typical fallback mass (several $10^{-3}\,M_\odot$) found in explosion models 
of SN~1987A progenitors \citep{Sukhbold:2016,Ertl:2019}, so a 
light-element-dominated envelope at early times is certainly plausible.

\section{Survival of Light Elements in the Envelope}
\label{Sec:burning}

\begin{figure}[t]
\includegraphics[width=\columnwidth]{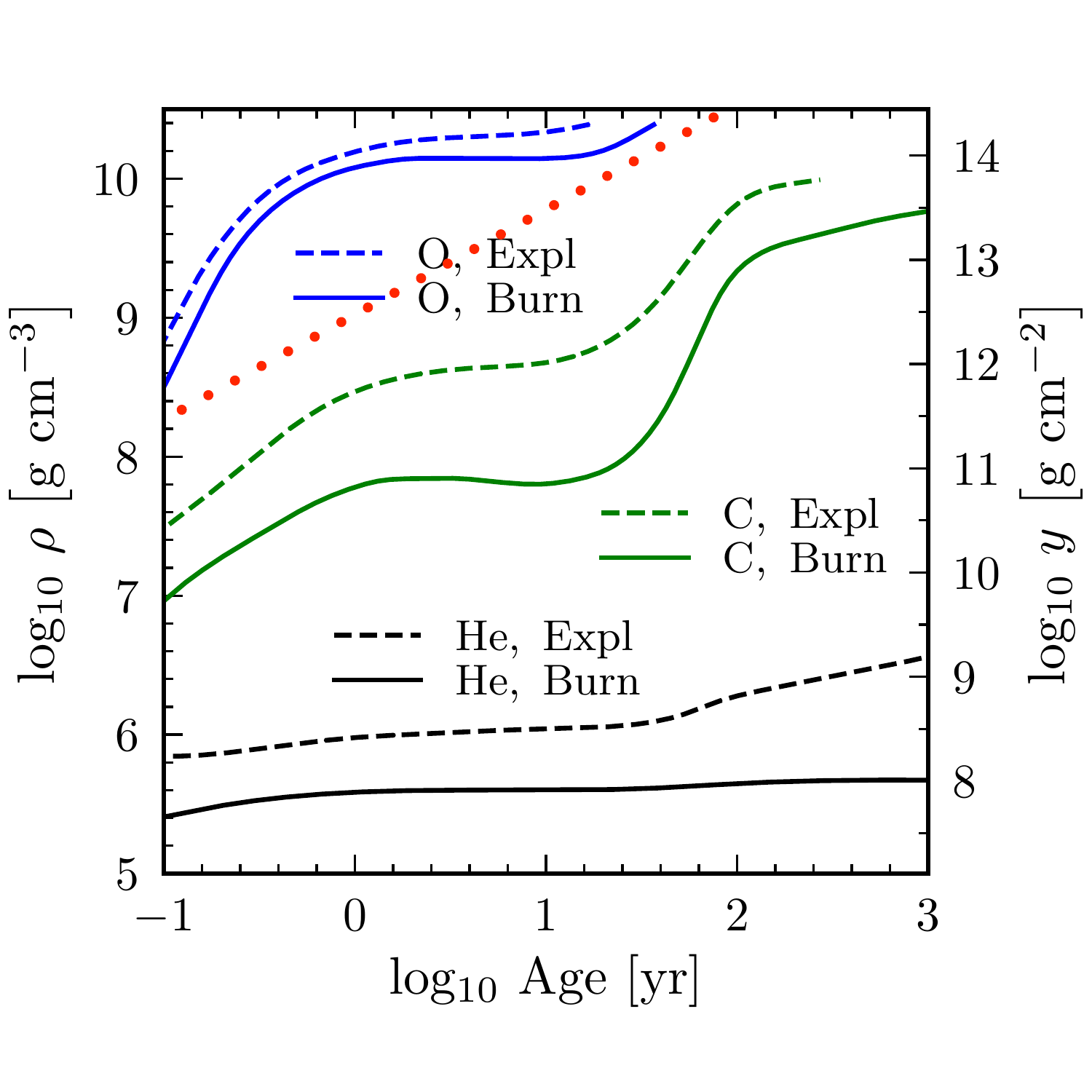}
\caption{Survival density of light elements (He, C, and/or O) versus time of deposition. 
Dashed curves show the density where the explosion criterion [\eq{Eq:Expl}] is met, whereas the continuous lines show the density where the burning timescale exceeds the star's age [\eq{Eq:Life}].
The diagonal dotted (red) line shows the maximum density reachable with accretion at the Eddington rate [\eq{Eq:Edd}].  
These thresholds depend on the star's $M$ and $R$ and can vary by factors of 
2--3; $M=1.4 \, M_\odot$ and the APR EOS have been assumed.}
\label{Fig:Nucl}
\end{figure}

The crust of a neo-neutron star, originally formed of beta-equilibrated (catalyzed) matter, may be altered by either fall-back soon after the supernova explosion or later accretion.
We will not discuss these scenarios here, but simply analyze the fate of light elements in the envelope of a young neutron star.
At temperatures well above $10^7$ K, they will be slowly depleted by thermonuclear fusion, or, if present at high-enough densities, can even burn explosively.

A simple criterion to determine the depth at which the burning becomes unstable is given by 
\be
\frac{d \epsilon_\mathrm{nucl}}{dT} > \left| \frac{d \epsilon_\mathrm{cool}}{dT} \right| \,,
\label{Eq:Expl}
\ee
where $\epsilon_\mathrm{nucl}$ is the thermonuclear energy generation rate and $\epsilon_\mathrm{cool}$ is the ``cooling rate" determined by the temperature gradient in the burning layer \citep{Bildsten:1998aa}.
The quantities $d \epsilon_\mathrm{cool}/dT$ and $d \epsilon_\mathrm{nucl}/dT$ are taken directly from cooling simulations and from the {\tt MESA} code \citep{Paxton:2011aa}, respectively.
A thermonuclear runaway results in an X-ray burst when the above criterion is satisfied and light elements will be processed into iron peak nuclei except in the outermost layers with column densities below $y \sim 10^6~\gcs$ (corresponding to densities below $10^4\gcc$)
that remain cold enough for H/He to survive.  At densities higher than established by Eq. (\ref{Eq:Expl}) the envelope must consist of heavy elements.

The second criterion for light element survival is that the burning timescale of a given nucleus, $\tau_\mathrm{nuc}$, be longer than the present age of the star, i.e.,
\be
\tau_\mathrm{nuc} \equiv \frac{n_\mathrm{nuc}}{r_\mathrm{nuc}} > \mathrm{Age} \,,
\label{Eq:Life}
\ee
where $r_\mathrm{nuc}$ is the burning rate and $n_\mathrm{nuc}$ the number density of the nucleus.
The burning of He results in C/O,  whereas C/O burning produces a blend of 
elements with $A \sim 24$--$32$, that, in terms of envelope structure and the 
resulting $T_\mathrm{b}$-$T_\mathrm{s}$ relationship, can be considered to 
be heavy elements.

\begin{figure*}
\centerline{\includegraphics[width=1.03\textwidth]{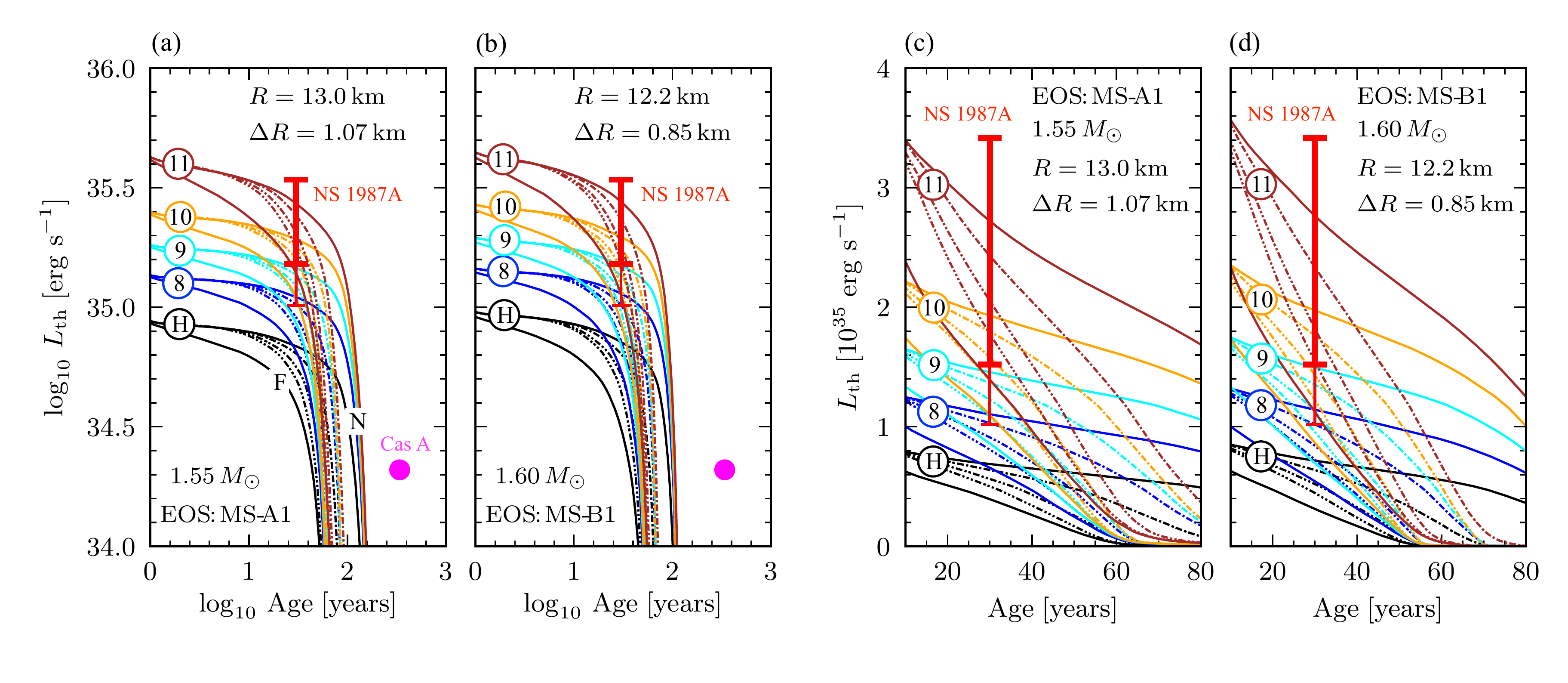}}
\caption{Thermal evolutions of relatively massive stars that have a direct Urca process acting in their inner core, with logarithmic (panels (a) and (b)) or linear (panels (c) and (d)) axes.
Each panel has five families of curves reflecting different envelope chemical compositions; each family has five member curves with different assumptions about crust n-superfluidity.
The notation of the curves and error bar follows that of \fig{Fig:Cool_SF_CasA}.
The  magenta dot shows $\Lth$ of Cas A.}
\label{Fig:Cool_DU}
\end{figure*}

A practical evaluation of these two survival criteria is shown in \fig{Fig:Nucl} using the temperature profile in a neo-neutron star obtained by \citet{Beznogov:2020aa}.  
Two important results of this inquiry are that 
1) light elements will be exhausted before they can produce a thermonuclear runaway, and 
2) O can survive at much higher densities than C due to its higher electric charge.

Optimal agreement between cooling models and the estimated luminosity of NS~1987A would imply that a thick\footnote{The thickness of the light element layer, from the photosphere down to $\rho_{\cal L}$, depends on the temperature, hence on the star's age, and the surface gravity $g_\mathrm{s}$. 
At the present age and crust temperature of NS~1987A, this thickness is 50, 
100, 225, and 450 meters for ${\cal L} = 8$, 9, 10, and 11, respectively, when 
$g_\mathrm{s} = 2\times 10^{14}$ cm s$^{-2}$. It scales approximately as 
$g_\mathrm{s}^{-1}$.}
layer of light elements, up  to densities  $\simge10^9 \gcc$, is present in the envelope and should consist of O since lighter elements cannot survive so deep.
The dotted line in \fig{Fig:Nucl} shows the maximum density that can be reached from Eddington-rate accretion over the life of the neutron star, determined by
\begin{equation}
    \frac{y_\mathrm{max}}{\mathrm{Age}} =  \dot{y}_\mathrm{Edd} \simeq 10^5\gcs{\rm~s}^{-1}
    \label{Eq:Edd}
\end{equation}
with $y$ the accreted column density.
A layer of oxygen reaching $10^9\gcc$ could be accumulated by accretion at the Eddington limit for a year or a few years, remaining unnoticed as long as radioactivity brightened the supernova ejecta. 
Even higher oxygen densities may be possible by super-Eddington accretion from a fallback disk or through neutrino cooling in accretion flows shortly after the birth of the neutron star.
We defer a more exhaustive study of the origin and survival of the light-element envelope to future work.

\section{Direct Urca Cooling}
\label{Sec:DUrca}

In \sect{Sec:minimal}, the evolution of \Lth was restricted to standard cooling. 
In high-mass neutron stars, however, it becomes more likely that the direct Urca (DU) or another enhanced cooling process can operate (see \app{app:DU}).
Enhanced neutrino cooling in the inner core leads to a much greater and more rapid drop of $T_\mathrm{s}$ and a shortening of the early plateau \citep{Page:1990aa,Page:1992aa}. 
Diffusion of the crustal thermal energy into the core determines the age [\eq{Eq:tau}] at which the crust and core reach thermal equilibrium after which $T_\mathrm{s}$ mirrors the core temperature.
This age is strongly dependent on the star's crust thickness, determined by the star's mass $M$ and radius $R$, and, to a lesser extent, the neutron superfluidity in the inner crust \citep{Lattimer:1994aa,Gnedin:2001aa,2008AstL...34..675S}.

\fig{Fig:Cool_DU} contains sets of DU cooling models with different assumptions about the core's EOS, the inner crust neutron superfluidity, and the composition of the envelope, analogous to standard cooling models illustrated in \fig{Fig:Cool_SF_CasA}.  
However, \fig{Fig:Cool_DU} focuses on relatively massive stars within the 
(1.22--1.62)\,$M_\odot$ range thought to be associated with NS 1987A for 
which the DU process can operate.  
The cases shown obey the constraint $R_{1.4}<13.5$ km employed earlier.
The EOSs MS-C1 and APR allow the direct Urca only for masses well above the estimated mass range for NS~1987A and are hence not included in \fig{Fig:Cool_DU} (see \app{app:tech}).

The age of NS~1987A is apparently too short to guarantee that operation of the DU process has resulted in $\Lth\ll\Lcs$.  
Only an age larger than about 100 years would be definitive in that regard.  
If the abundance of light elements in the envelope is high enough and the crust is not completely superfluid, it is possible that $\Lth=\Lcs$ at the present time.  
Note that these are the same conditions on the crust and envelope that are required for $\Lth=\Lcs$ in the case of standard core cooling.
Thus, it is not possible at present to discriminate between these two scenarios.

Nevertheless, an exciting possibility is that enhanced cooling could become apparent on a time scale as short as a few years, in the event that $\Lth\sim\Lcs$, through the predicted dimming of the dust blob.  
If this dimming is not eventually observed, one could infer that either the DU process is not operating or the observed blob heating is due to pulsar spin-down or accretion.  
In the future, as the present upper limit to the power of a pulsar in the remnant 
of SN~1987A, which is now only about 1.5--3.5\,\Lcs, decreases, the 
likelihood of a thermal source becomes even greater than it is now, and 
conclusions concerning the DU process in NS~1987A will become firmer.

In fact, this could happen for models permitting enhanced cooling with exotica (see \app{app:DU}) with even smaller masses than those permitting the nucleon DU process.  
Nevertheless, an important constraint on such models is the apparent agreement of the Minimal Cooling paradigm \citep{Page:2004fr} with the observed thermal emissions of older neutron stars deduced from their temperatures and ages, suggesting enhanced cooling of any type exists only in relatively massive neutron stars. 

\section{Comparison of the neutron stars  in SN~1987A and Cas~A} 
\label{Sec:Comps}

It is interesting to compare NS 1987A with Cas~A, the CCO in the Cassiopeia A SNR and the second-youngest-known neutron star, about 340 years old.
Its relatively high $T_\mathrm{s}$ gives no evidence for DU cooling.
Theoretical models (see panels (d) through (f) of \fig{Fig:Cool_SF_CasA}) capable of explaining both observed values of its current luminosity \Lth and its recently detected \citep{Heinke:2010aa} rapid cooling ({\it i.e.,} large $|d\Lth/dt|$) invoke Cooper-pair neutrino emission due to the
onset of neutron superfluidity in the core \citep{Page:2011aa,Shternin:2011aa} and pointedly prohibit DU cooling.
Although the n-$^3$P$_2$ and n- and p-$^1$S$_0$ gaps were adjusted in 
these panels to allow the ${\cal L} =8$ models to satisfy the Cas A constraints 
for $1.4\,M_\odot$, all light-element envelope models could also be made to fit 
with minor gap changes. 
On the other hand, none of the heavy-element models (H) can satisfy the Cas A constraints.

 \citet{2013ApJ...779..186P} and \citet{2018ApJ...864..135P} have argued that the cooling rate of Cas A is much smaller than originally reported (but see \citealt{Wijngaarden:2019aa}).  In this case, the constraint on the size of the n-$^3$P$_2$ gap is modified or there could even be no constraint at all.
This, however, does not spoil the agreement of Cas A's $L_\mathrm{th}$ with the same class of models we present for NS 1987A
and, hence, does not alter our conclusions.

It is known that Cas~A was a Type IIb event \citep{Krause:2008ep}, which 
means that its progenitor had lost nearly all of its hydrogen, retaining less than 
0.1\,$M_\odot$ before it collapsed and exploded. 
This also means that there was no reverse shock from the supernova shock propagating through the He/H interface.  
\cite{Orlando:2016} determined by detailed modeling of the remnant evolution 
an ejecta mass of about $4\,M_\odot$.  
Adding in the neutron star's baryon mass suggests a mass of 
$(5.5$--$6)\,M_\odot$ for the He-core mass of the progenitor. 
From Fig. 1 in \cite{Woosley:2019}, one finds a progenitor zero-age main 
sequence mass in the range of $(18$--$20)\,M_\odot$, possibly only slightly 
more massive than the progenitor of SN 1987A.  

Given similar progenitors, it is therefore tempting to examine theoretical cooling models of young neutron stars that fit both NS 1987A and Cas A.  
This apparently requires their envelopes to have relatively abundant light elements at age 30 years and the same or possibly lower abundances by age 340 years. 
This could happen if the envelope light-element mass decreased with time as larger and larger fractions of their envelopes catalyze. 
This inferred evolution of the envelope composition would be extremely interesting.

An alternative, of course, is that  these two neutron stars were born with significantly different light-element compositions that didn't evolve.
Despite similar progenitor masses, there still could be considerable differences in resulting neutron star masses and supernova explosion properties ({\it e.g.}, see the considerable case-to-case variations in Fig. 14 of \citealt{Ertl:2019}).  
Indeed, the explosion energy of the Cas A supernova is estimated to have been 
(2--2.5) bethe, considerably greater than that inferred for SN 1987A (about 
1.2--1.5 bethe).
Therefore, and because the progenitor had stripped nearly all of its hydrogen envelope, the fallback mass for Cas A is likely to have been less than for SN1987A, suggesting that NS 1987A's envelope was born with more light elements. 
Unlike the case for the envelope composition, no particular model of the n-$^1$S$_0$ gap is favored  for Cas A.  
However, since the gap model cannot evolve, models that fit both Cas A and NS 1987A observations favor smaller predicted values for $T_\mathrm{c}(\rho)$ close to the neutron drip point.

Finally, while the evidence is strong that the DU process (or other enhanced cooling process) is not operating in Cas A, the evidence in the case of NS 1987A is not conclusive at this time.  
Nevertheless,  given that small mass differences do not appreciably affect the cooling curves in \fig{Fig:Cool_SF_CasA}, both Cas A and NS 1987A are compatible with the Minimal Cooling scenario ({\it i.e.}, with no enhanced cooling.)

In summary, the fact that both NS 1987A and Cas A match a variety of cooling curves with similar assumptions about the envelope composition and the n-$^1$S$_0$ superfluid gap within the Minimal Cooling paradigm lends further support to the hypothesis that the observed excess dust blob luminosity in SN 1987 A is due to thermal power from the cooling of NS 1987A.

\section{Discussion and Conclusions} 
\label{Sec:Disc}

The luminosity \Lcs $ \sim (1.5$--$3.5) \times 10^{35} \; \ergs$ observed from 
a dust blob at the expected position of the compact remnant from SN~1987A 
appears to originate from an embedded or nearby {source} rather than from 
more widely distributed radioactivities, {\it e.g.,} from $^{44}$Ti.  
Hydrodynamic models of core-collapse supernovae compatible with the inferred SN~1987A progenitor mass and radius, the supernova light curve, the observed neutrinos, the properties of radioactive-decay lines, and the observed amounts and distribution of ejected heavy elements in the expanding SNR, also support a compact remnant mass much smaller than $M_{\rm max}$, so that a neutron star rather than a black hole is likely to have formed.  
Observations have already set an upper limit to the power of a hidden source in the SN~1987A remnant that is less than about 3.5 times the dust blob's luminosity.
In all likelihood, if the excess emission from the hot dust blob is due to a neutron star, its inferred power stems from pulsar spin-down, accretion, or from surface thermal emission, and the source is either embedded in the blob or located nearby. 

If the source is a pulsar, its period $P$ and spin down rate $\dot P$ (or, equivalently, polar magnetic field strength $B_p$) must have values that fall between the solid red lines in Fig.~\ref{Fig:P_Pdot} or slightly above them.  
The required power, which must be within the range 
$26\,L_\odot$--$138\,L_\odot$, is much less than those of the young Crab 
($W\sim10^5\,L_\odot$) and Vela ($W\sim1800\,L_\odot$) pulsars.
Furthermore, the factor 5 of this range is extremely small compared to the 
possible range (3$\cdot10^{-4}\,L_\odot$--3$\cdot10^4\,L_\odot$)  observed 
among other known regular, non-recycled pulsars.
Therefore $W\propto B_p^2/P^4$  inferred for a pulsar in the SN~1987A remnant requires  a major degree of fine-tuning, strongly disfavoring this scenario.
A similar fine-tuning argument can be made in the case of accretion, either by a neutron star or a black hole. But if one or the other of  spindown and accretion is the correct source, then it is possible that NS~1987A has a pure heavy-element envelope or, alternatively, is undergoing enhanced cooling, because thermal cooling would then contribute to $L_\mathrm{cs}$ only on a subdominant level.

\begin{table*}[ht]
\caption{Scenario Summary [Required power range taken from Fig. \ref{Fig:Cool_SF_CasA}.]}
\begin{center}
\begin{tabular}{|c|c|c|c|}
\hline
Scenario/source&Expected power range&Advantages&Disadvantages\\
\hline
$^{44}$Ti&$L_\mathrm{Ti}\lesssim13L_\odot$ with radio-&$^{44}$Ti is observed in the&The required blob mass is too large;\\[-3pt]
heating&active heating to 22\,K.&remnant of SN~1987A.&the blob is also optically thin to $\gamma$-rays.\\
\hline&&&\\[-12pt]\hline
& &Blob's location offset from progen-&\\[-3pt]
compact & &itor's and matches the kick velocity&\\[-3pt]
object:& &predictions from asymmetrical&\\[-3pt]
neutron star& &distributions of $^{56}$Co and $^{44}$Ti.&\\[-3pt]
or black hole& &This kick speed agrees with&\\[-3pt]
& &the known pulsar distribution.&\\
\hline&&&\\[-12pt]\hline
\multirow{5}{*}{neutron star}& &SN~1987A simulations imply&\\[-3pt]
 & &$1.22\,M_\odot\le M\le  1.62\,M_\odot$.&\\[-1pt]
& &The SN~1987A $\nu$ signal implies&\\[-3pt]
 & &$0.98\,M_\odot\le M\le 1.81\,M_\odot$.&\\[-1pt]
  & &Both imply $M<M_\mathrm{max}$.  &\\[-8pt]
\hdashrule[.2ex]{2.35cm}{1pt}{0.25cm}\\[-5pt]
\cline{2-4}
&\multirow{7}{*}{$5.8\,L_\odot-72\,L_\odot$}&$L_\mathrm{th}>26\,L_\odot$
needs little fine-tuning.&\\[-1pt]
&&Evolution consistent with Cas~A.&\\[-1pt]
&& Possible light-element envelope is & $\Lth$ is likely insufficient unless \\[-3pt]
thermal&& compatible with surrounding dust & the neutron star is inside the blob. \\[-3pt]
emission&& and CO, SiO molecules; survival of &\\[-3pt]
&&this envelope supported by theory.&\\[-1pt]
&& Decent chance SNRs have CCOs.&\\[-8pt]
\hdashrule[.2ex]{2.35cm}{1pt}{0.15cm}\\[-5pt]\cline{2-4}
&\multirow{4}{*}{3$\cdot10^{-3}\,L_\odot-$3$\cdot10^5\,L_\odot$} &The 
pulsar does not have to &$W<138\,L_\odot$ from SNR observations, so\\[-3pt]
pulsar&& be embedded in the blob. &the pulsar must be close to the blob.  $W$\\[-3pt]
spindown&& $W>26\,L_\odot$ is possible. &and pulsar location require 
fine-tuning.\\[-1pt]
& & &$B_p/P^2$ also requires fine-tuning.\\[-8pt]
\hdashrule[.2ex]{2.35cm}{1pt}{0.15cm}\\[-5pt]
\cline{2-4}
accretion&$\lesssim\,$3.4$\cdot10^5(M/M_\odot)\,L_\odot$&$L_\mathrm{acc}>26\,L_\odot$
 is possible. &$L_\mathrm{acc}$ requires fine-tuning.\\
\hline&&&\\[-12pt]\hline
\multirow{5}{*}{black hole}&& &SN~1987A progenitors have small\\[-3pt]
&&&core masses; also, SN~1987A's\\[-3pt]
&&&observed explosion energy was high,\\[-3pt]
&&&implying a small fallback mass.\\[-3pt]
&&& Both strongly suggest $M<M_\mathrm{max}$. \\[-8pt]
\hdashrule[.2ex]{2.35cm}{1pt}{0.25cm}\\[-5pt]
\cline{2-4}
\multirow{1}{*}{accretion}&\multirow{1}{*}{$\lesssim\,$3.4$\cdot10^5(M/M_\odot)\,L_\odot$}&$L_\mathrm{acc}>26\,L_\odot$
 is possible.&$L_\mathrm{acc}$ requires fine-tuning.\\
\hline
\end{tabular}
\label{tab:table}
\end{center}
\end{table*}

A more likely possibility, favored strongly by the similar magnitudes of \Lcs and the expected \Lth for a $\sim$30 year old remnant, is the scenario that the dust blob is powered by the thermal surface luminosity of a neutron star. If correct, a long heat transport timescale in the crust and a large effective stellar temperature are favored, implying relatively limited crustal n-$^1$S$_0$ superfluidity near the neutron drip point and an envelope with a thick layer of light elements, respectively.
At this time, it is not possible to definitely rule out the occurrence of an 
enhanced neutrino cooling process, although nucleon DU cooling is disfavored 
because the star's mass is expected to be smaller than about $1.62\,M_\odot$ 
and the observed properties of most other cooling neutron stars {are} 
consistent with the Minimal Cooling paradigm, {which} specifically disallows 
the operation of the DU process.

This is particularly true of the second youngest-known neutron star, Cas~A, and, in fact, we have shown that it is straightforward to understand both objects with common standard cooling models as long as they both have light-element envelopes.   
In support of this thesis, we demonstrated that nuclear burning rates are small enough to allow accreted light elements to survive to reasonably high densities despite the high temperatures in their early evolution. 

The mass of light elements in the envelope needed for optimal agreement with 
observations for NS 1987A and Cas A seems to be relatively large, with 
$\rho_{\cal L}\simge10^9\gcc$, but this could be somewhat relaxed if the 
neutron star mass is at the high end of expectations, around $1.6\,M_\odot$.
There is some evidence from the comparison of NS~1987A and Cas~A that the masses of light elements in their envelopes have decreased with time, but it could also be that these objects were born with different envelope compositions.
While NS~1987A favors models with relatively small  n-superfluid critical temperatures in the crust,  Cas A is agnostic to these details. 

Interestingly, if NS~1987A does have enhanced cooling, its \Lth is predicted to decrease quickly enough to be observed in the excess infrared dust emission on timescales as short as a few decades, if not years. 
Of course, if such a decrease in the dust blob's emission is not seen, or if thermal X-rays from the cooling neutron star eventually materialize,  the evidence would be strong that the DU process is not operating in NS~1987A. 
The predicted mass of NS~1987A is (1.22--1.62)\,$M_\odot$, well below 
$M_\mathrm{max}$, with central densities in the vicinity of $(8$--$10)\times 
10^{14}\gcc$ or (3--4)$\rho_\mathrm{s}$,
$\rho_\mathrm{s} \simeq 2.8\times 10^{14} \gcc$ being the nuclear saturation density.
A conclusion regarding the operation of a DU process would have important ramifications for the behavior of the nuclear symmetry energy in the same density range.  
This information can be combined with observational inferences of neutron star radii, which are mostly sensitive to the density dependence of the nuclear symmetry energy around $2\rho_\mathrm{s}$, to form a more complete understanding of the full density dependence of the symmetry energy.

While we tend to favor a thermally cooling CCO heating the blob, because its emission can naturally match the inferred luminosity of the blob, alternative possibilities such as a pulsar or a compact object accreting at a low rate of $\sim$($10^{-11}$--$10^{-10})\,M_\odot$\,yr$^{-1}$  \citep{Alp:2018ao} cannot be excluded.\footnote{We used the condition $L_\mathrm{acc} = L_\mathrm{cs}$ with $L_\mathrm{acc} = \eta\,\dot{M}\,c^2 = 148\,L_\odot\,\eta\,[\dot{M}/(10^{-11}\,M_\odot/\mathrm{yr})]$, where $\eta \sim$\,0.1--0.4 is the energy conversion efficiency parameter.}
 
As discussed in the Introduction, the observed location of the blob north of the SN center is consistent with expectations for the position of a compact remnant, based on a comparison of 3D explosion models and the distribution of [SiI]+[FeII] in HST and VLT images~\citep{Janka:2017aa}.
Furthermore, the kick speed (300--500 \,km\,s$^{-1}$) and direction inferred from the displacement of the blob center from the progenitor's location (about 30$^\circ$ between the kick vector and observer direction) are consistent with predictions of 3D models that are compatible with the red-shifted $^{44}$Ti decay line and the $^{56}$Co-decay line profile observed for SN~1987A \citep{Jerkstrand2020}. 
Our theoretical analysis of the neutron star  thermal cooling was also 
constrained by the mass range  (about 1.22--1.62\,$M_\odot$) suggested by 
current explosion simulations of SN~1987A, which are based on proposed 
progenitor models that possess explosion properties consistent with 
observations. Further predictions by 3D explosion modeling, in particular of 
the neutron star  spin, are hampered by our incomplete knowledge of the 
progenitor's core rotation, a lack of observational constraints, and the influence 
of stochastic hydrodynamic processes during the explosion. Moreover, 
magnetohydrodynamic models are not advanced enough to yield predictions of 
the magnetic fields of young neutron stars.

Some arguments for and against these various scenarios are summarized in Table \ref{tab:table}.

The unique situation of having first indications of a compact remnant in SN~1987A warrants follow-up observations to monitor the infrared emission from the dust blob, improve the upper limit to the power of a hidden pulsar, detect or refine upper limits to X-ray thermal emission from a young neutron star, or identify pulsed emission. The periodic variations of pulsar radiation, the cooling evolution of the thermal emission of a CCO, or irregular luminosity variations of an accretion source would provide clues  about the exact nature of the object that heats the dust blob.  Observational constraints  on the mass and density of the dust blob may also help to estimate its transparency to X-rays. A better determination of the blob temperature and of its surroundings would help to get tighter limits on the blob luminosity, and direct measurements of radioactive $\gamma$-rays from $^{44}$Ti decay with high spatial resolution, as achieved with NuSTAR in the case of the fifteen-times closer Cas~A, would clarify the role of radioactive heating in the blob.


\pagebreak
{\bf Note added in proof:}
We note that \citet{Orlando:2020aa}, based on 3D hydrodynamic modeling
of SN~1987A for different progenitors and explosion geometries, confirmed
the conclusions of \citet{Janka:2017aa} for the neutron star kick direction
and magnitude. They also found their favored model to be consistent with
an association between the neutron star and the dust blob in the ALMA images,
which is slightly offset to the north-east direction from the estimated
position of the progenitor star of SN~1987A \citep{Cigan:2019aa}.

\newpage
\acknowledgments
We are grateful to Frank Timmes for helping us to implement the MESA nuclear network in our investigation. We thank Sanjay Reddy for useful comments.  DP and MB acknowledge financial support by the Mexican Consejo Nacional de Ciencia y Tecnolog{\'\i}a with a CB-2014-1 grant $\#$240512 and the Universidad Nacional Aut\'onoma de M\'exico through an UNAM-PAPIIT grant \#IN109520.
MB also acknowledges support from a postdoctoral fellowship from UNAM-DGAPA.
IG is partially supported by a SNI-Conacyt ``Ayudante de Investigador'' fellowship.
The researches of JML and MP are supported by the U.S. Department of Energy Grant Nos. DE-AC02-87ER40317 and 
DE-FG02-93ER40756, respectively. 
HTJ acknowledges funding by the European Research Council through Grant ERC-AdG No.~341157-COCO2CASA
and by the Deutsche Forschungsgemeinschaft (DFG, German Research Foundation)
through Sonderforschungsbereich (Collaborative Research Centre)
SFB-1258 ``Neutrinos and Dark Matter in Astro- and Particle Physics
(NDM)'' and under Germany's Excellence Strategy through
Cluster of Excellence ORIGINS (EXC-2094)---390783311.


\clearpage
\appendix

\section{Details of the EOS models}
\label{app:tech}

\begin{table*}[b]
	\caption{RMFT coupling strengths. Values of the meson masses used are 
	$m_\sigma = 660$ MeV, $m_\omega = 783$ MeV, and $m_\rho = 770$ 
	MeV.
		All couplings are dimensionless except for $\kappa$ whose unit is 
		fm$^{-1}$.
		{See \citet{Han:2019aa} for notations.}}
	\begin{center}
		\begin{tabular}{cccccccccc}
			\hline
			Models & $g_\sigma$ & $g_\omega$ & $g_\rho$ & $\kappa$ & 
			$\lambda$ & $\zeta$ & $\xi$ & $\Lambda_\sigma$ & 
			$\Lambda_\omega$ \\
			\hline
			MS-A1 &   $11.833$   &  $10.702$ &   $8.578$ & $4.374\times 
			10^{-2}$ & $-2.829\times 10^{-2}$  & $4.767\times 10^{-4}$  &  
			$1.307$    &  $7.443\times 10^{-3}$  & $6.545\times 10^{-3}$ \\
			MS-B1 &   $11.359$   &  $10.014$ &   $8.486$ &  $5.709\times 
			10^{-2}$ & $-3.901\times 10^{-2}$  & $3.848\times 10^{-4}$  &  
			$1.103$   &  $9.599\times 10^{-3}$  & $1.202\times 10^{-2}$ \\
			MS-C1 &   $11.412$   &  $ 9.945$  &   $9.667$ & $6.521\times 
			10^{-2} $ & $-4.907\times 10^{-2}$  & $9.555\times 10^{-4}$  &  
			$0.3462$ &  $9.652\times 10^{-3}$  & $4.019\times 10^{-2}$ \\
			\hline
		\end{tabular}
	\end{center}
	\label{Tab:EOS_coupling}
\end{table*}

\begin{table*}[b]
	\caption{Physical and astrophysical properties of the four EOSs used. Listed 
	physical properties are:
		saturation density $n_0$, binding energy $B$, compression modulus 
		$K_0$, symmetry energy $S_2$ and its slope parameter $L$,
		Landau effective mass $m_\mathrm{L}^*$ (normalized to the nucleon 
		mass 938 MeV).
		The astrophysical properties are the maximum mass $M_\mathrm{max}$, 
		radius of a $1.4 \,M_\odot$ star $R_{1.4}$, and the threshold stellar mass
		$M_\mathrm{DU}$ and baryon density $n_\mathrm{DU}$ for the onset 
		of the nucleon direct Urca process.}
	\begin{center}
		\begin{tabular}{ccccccccccc}
			\hline
			Models &   $n_0$     &  $B$  & $K_0$ & $S_2$ & $L$ & 
			$m_\mathrm{L}^*$ & $M_\mathrm{max}$ & $R_{1.4}$ & 
			$M_\mathrm{DU}$ & $n_\mathrm{DU}$ \\
			& fm$^{-3}$ &  MeV &    MeV  &  MeV  &  MeV  
			&                                 &     $M_\odot$           &     km       &   
			$M_\odot$             &       fm$^{-3}$         \\
			\hline 
			MS-A1 &     0.149    & 16.0  &   234     &   30.6  &  73.9  &          
			0.704             &            2.288            &    13.2      &        
			1.50                   &         0.410              \\
			MS-B1 &     0.159    & 16.1  &   222     &   31.0  &  71.2  &          
			0.717             &            2.155            &    12.5      &        
			1.55                   &         0.498               \\
			MS-C1 &     0.158    & 16.0  &   201     &   30.9  &  56.0  &          
			0.722             &            2.118            &    12.0      &        
			1.90                   &         0.686               \\
			APR    &      0.160    & 16.0  &  266      &   32.6  &  58.5  &          
			0.698             &            2.183            &    11.6      &        
			1.97                   &         0.774               \\
			\hline
		\end{tabular}
	\end{center}
	\label{Tab:EOS_prop}
\end{table*}

In this work, we have selected a set of dense matter EOSs that provide neutron 
star models with radii at $1.4 \, M_\odot$ between 11.5 and 13.5 km and a 
maximum mass above $2 \,M_\odot$.
This range of radii is deduced from constraints on the nuclear symmetry energy \citep{Lattimer13} and on the joint analysis \citep{Raaijmakers20} of LIGO/Virgo data for GW170817 \citep{De18, Abbott18} and NICER data for PSR J0030+0451 \citep{Riley19,Raaijmakers19, Miller19}.
Selecting EOSs with maximum mass $M_{\mathrm{max}}\simge 2\, M_\odot$ 
is required by the existence of three pulsars with measured mass around or 
above $2\, M_\odot$:
PSR J1614-2230 \citep{Demorest:2010aa}, PSR J0348+0432 \citep{Antoniadis:2013aa}, and PSR J0740+6620 \citep{Cromartie:2020aa}).
The well-known APR EOS \citep{Akmal:1998aa} is taken as a first reference EOS.  A second, stiffer, reference EOS is afforded by a relativistic mean-field theory (RMFT) scheme \citep{Mueller:1996pm}, as presented in \citet{Han:2019aa}, with three sets of coupling constants listed in \tab{Tab:EOS_coupling}.

\begin{figure*}
\centerline{\includegraphics[width=0.9\textwidth]{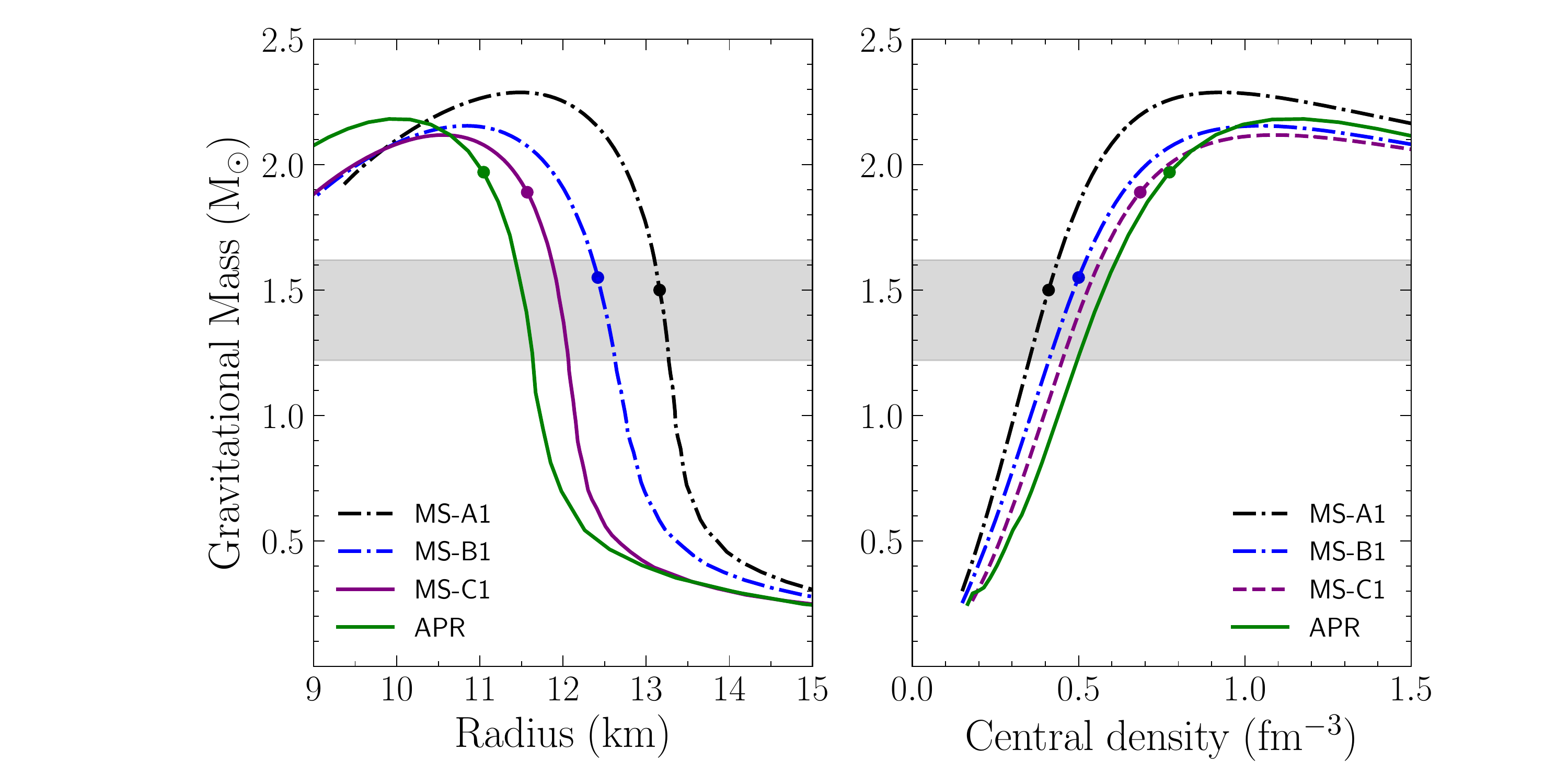}}
\caption{Mass versus radius (left panel) and central density (right panel) for the EOSs used in this work.
Dots on the curves show nucleon direct Urca thresholds. 
The shaded band shows the expected mass range for the compact object in SN 1987A.
\label{Fig:EOS_MRn}}
\end{figure*}

\tab{Tab:EOS_prop} compares the main physical and astrophysical properties resulting from our chosen EOSs.
In \fig{Fig:EOS_MRn}, we present the mass-radius and mass-central density curves for these EOSs.
In this figure, the gray band indicates the  most likely mass range of 
(1.22--1.62)\,$M_\odot$ (see discussion around Eq.~(\ref{eq:BGmasses})) for 
the compact object in SN~1987A, as suggested by explosion simulations of 
SN 1987A  progenitors  \citep{Utrobin:2019,Ertl:2019}.  These supernova 
simulations did not follow neutron star  formation and cooling in detail, but 
were focused on producing explosions (in 1D and 3D) with the energy and the 
$^{56}$Ni  yield observed for SN 1987A. 
These observational constraints determined the mass cut between the supernova ejecta and the neutron star  in the simulations, and thus the neutron star  mass but not its radius.\footnote{Notice that the proto-neutron star  radii in supernova simulations are determined by a finite-temperature EOS that may predict different radii from the cold EOSs used in this work. However, the cold counterpart of the EOS used in those simulations yields a mass-radius curve similar to the ones 
 in \fig{Fig:EOS_MRn}.}

Despite an extensive search, we could not find an EOS resulting in $R_{1.4}$ 
below 12 km and $M_\mathrm{DU}$ below $1.9\, M_\odot$ within the RMFT 
scheme employed.  
An additional criterion for selecting the EOSs MS-A1 and MS-B1 was that 
they also allow the occurrence of the nucleon direct Urca process at masses 
below the estimated upper limit of the mass of NS 1987A, i.e., 
$1.62\,M_\odot$.  
Although values of $R_{1.4}$ and $M_\mathrm{max}$ for MS-C1 are
consistent with current constraints, the nucleon DU process is
permitted only for stars with $M \geq 1.9\,M_\odot$, similar to APR for
which $M$ needs to be $\geq 1.97\,M_\odot$.  As is discussed in
\app{app:DU}, models containing exotica, e.g., quarks, in the core
may permit alternative DU processes for smaller masses.

\section{Neutrino emission processes}
\label{app:DU}

 The simplest neutrino cooling process is the direct Urca (DU) process \citep{Boguta:1981aa,Lattimer:1991aa} involving nucleons
\begin{equation}
    n\rightarrow p+e^-+\bar\nu_e;\qquad  n\rightarrow n+e^++\nu_e.
    \label{eq:urca}
\end{equation}
Since neutrinos readily escape from the core, energy is lost in both halves of the cycle, leading to cooling.  
In completely degenerate matter, this process cannot occur because there are no available final nucleon momentum states. 
At finite temperatures, limited phase space is available near the Fermi surface.  
Beta equilibrium guarantees energy conservation for Eq. (\ref{eq:urca}), but momentum conservation requires 
participant Fermi momenta to fulfill $k_\mathrm{F}(p) + k_\mathrm{F}(e) \ge k_\mathrm{F}(n)$ 
equivalent in the absence of muons to the proton fraction condition $n_p/(n_n+n_p)\ge1/9$ (with muons, the minimum proton fraction $\simeq 0.14$).
Depending on the stellar mass and the behavior of the dense-matter nuclear symmetry energy, this condition may not be fulfilled anywhere and the DU process is forbidden.  
Since the proton fraction generally increases with density, however, the DU 
process may begin to operate in a sufficiently massive neutron star.
 
Similar DU processes may also exist involving hyperons, meson condensates or quark matter (see, e.g. \citet{Pethick:1992,Prakash:1994}).  
The possibility of a quark DU process operating in hybrid hadron-quark stars described by the models of \citet{Han:2019aa}, which  satisfy current laboratory and astrophysical constraints, will be reported elsewhere.
If no DU process operates, neutrino cooling is dominated by the modified Urca (MU) process \citep{Friman:1979} which requires a bystander nucleon in the initial and final states to soak up the excess momentum.  
In partially degenerate matter, the MU process is $(T/E_F)^2\sim10^{-6}$ times less efficient ($E_F\simge40$ MeV is the nucleon Fermi energy), resulting in substantially less-rapid cooling.  
The Minimal Cooling paradigm \citep{Page:2004fr} asserts that enhanced 
cooling does not generally occur, so that any star observed to be too cold for 
standard cooling to explain must have had enhanced cooling.

\clearpage
\section{Neutron Superfluidity in the Inner Crust}
\label{app:SF}

\begin{figure*}
	\centerline{\includegraphics[width=0.9\textwidth]{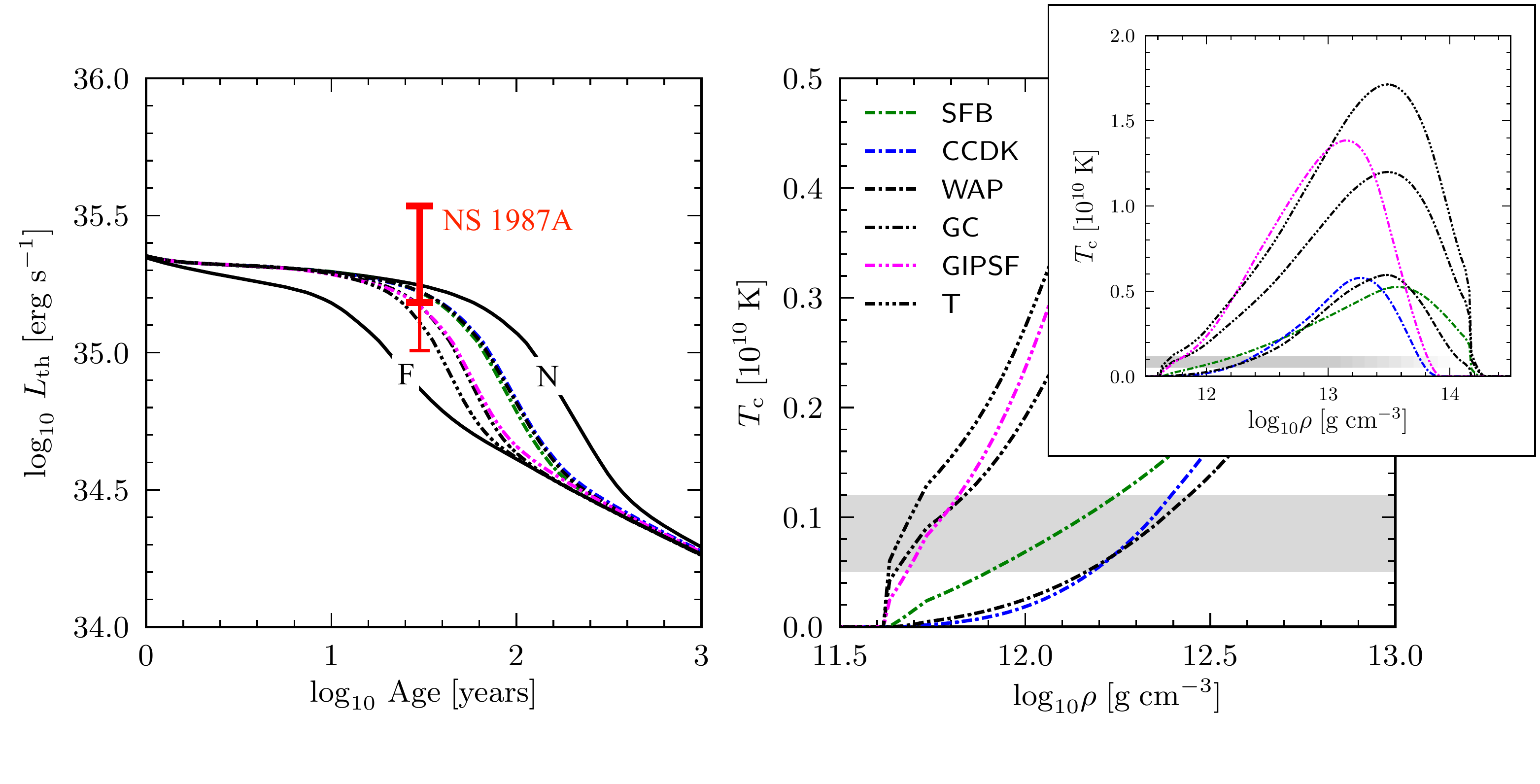}}
	\caption{The left figure shows evolutions of \Lth under $2+6$ different 
		assumptions about the size and density dependence of the inner crust 
		dripped neutrons $^1$S$_0$ superfluidity $T_\mathrm{c}(\rho)$ relation.
		The neutron star model (MS-C1) is the same as in panel a of 
		\fig{Fig:Cool_SF_CasA}, with $\mathrm{log}_{10} \, \rho_\mathrm{\cal 
			L} =$ 10.  Solid curves indicate the two extreme models F and N for 
		which all dripped neutrons are assumed to be either fully superfluid or not 
		superfluid at all.  The other six curves show results for the microscopic 
		models of $T_\mathrm{c}(\rho)$ in the inner crust displayed in the right 
		figures. The main right figure shows details of $T_\mathrm{c}(\rho)$ near 
		the neutron drip point and the inset shows their entire behaviors. The 
		models used are
		T from \citet{1984PThPh..71.1432T},
		GIPSF from \citet{2008PhRvL.101m2501G},
		GC from \citet{2008PhRvC..77c2801G},
		WAP from \citet{1993NuPhA.555..128W},
		CCDK from \citet{1993NuPhA.555...59C}, and
		SFB from \citet{2003NuPhA.713..191S}.
		The error bar is taken from Fig. \ref{Fig:Cool_SF_CasA} and the gray 
		shaded band indicates the range of temperatures encountered by the 
		cooling models in \fig{Fig:Cool_SF_CasA} for ages between 10 and 30 
		years.}
	\label{Fig:SF}
\end{figure*}

Dripped neutrons in the inner crust are predicted to form a superfluid and the 
results presented in this paper indicate that the future time evolution of the 
thermal luminosity \Lth of NS 1987A may offer some valuable information 
about this phenomenon.
We present here some details about the superfluidity models we employed, and 
refer the reader to \citet{Page:2014aa} for a detailed description, and their 
physical effects.

We considered six different microscopic neutron $^1$S$_0$ gap calculations, 
the same as those used in \citet{Page:2009dd}, and their corresponding 
$T_\mathrm{c}(\rho)$ curves are displayed in the right panel of \fig{Fig:SF}.
(The two extreme models F and N in which we arbitrarily imposed that all dripped neutrons in the inner crust are either fully paired, i.e., with $T_\mathrm{c}(\rho) \equiv 10^{10}$ K everywhere, or fully unpaired, i.e. with $T_\mathrm{c}(\rho) \equiv 0$ K everywhere, are not shown.)
The left panel of this figure shows the evolution of \Lth under these various pairing assumptions for the same cases as in panel (a) of \fig{Fig:Cool_SF_CasA} with $\mathrm{log}_{10} \rho_{\cal{L}} = 10$.

Apart from the two extreme F and N cases, the cooling curves in the left panel of \fig{Fig:SF} clearly separate in three  groups which, as seen in the right panel, correlate strongly with how rapidly $T_\mathrm{c}(\rho)$ grows with $\rho$ at low densities just above the neutron drip point (at $\rho_\mathrm{d} = 4.3 \times 10^{11} \gcc$).
A curious, and {\em a priori} not obvious, result is that the cooling curves have no significant sensitivity to the behavior of the gaps at high densities.
This is immediately seen by noticing that the evolutions of \Lth for the models SFB, CCDK, and WAP  are practically indistinguishable, despite the facts that these gaps, which are very similar at low densities, have very different high-density behaviors.
The same applies for the gaps GC and  GIPSF.  As a result, only results for the three gaps T, GC and WAP are presented in Figs. \ref{Fig:Cool_SF_CasA} and \ref{Fig:Cool_DU} and the line style in these two figures is the same as in \fig{Fig:SF}.
Finally, the maximum values reached by $T_\mathrm{c}(\rho)$ have little effect, being much larger that the crust temperature  during the early plateau (gray band in the right panel of the figure).

\clearpage


\end{document}